\documentclass[aps,epsfig,nofootinbib,a4paper,12pt,floatfix,amsfonts,amsfonts,amsmath]{revtex4}  

\usepackage[pdftex]{graphicx}
\usepackage[utf8]{inputenc}
\usepackage{lipsum}
\usepackage{clipboard}
\newclipboard{myclipboard}
\usepackage{color}
\usepackage[letterpaper]{geometry}
\usepackage{graphicx}
\usepackage{setspace}
\usepackage{natbib}
\usepackage{url}
\usepackage{bm}
\usepackage{textcase}
\usepackage[english]{babel}

\addto\captionsenglish{%
}

\usepackage{hyperref}
\usepackage{lmodern}
\usepackage[english]{babel}
\usepackage{float}
\usepackage{slashed}
\usepackage{amsmath}
\usepackage{epstopdf}
\usepackage{latexsym}
\usepackage{amssymb}
\DeclareMathOperator{\arctanh}{arctanh}
\usepackage{float}
\usepackage{slashed}
\usepackage{amsmath}
\usepackage{epstopdf}
\usepackage{latexsym}
\usepackage{amssymb}
\usepackage[center]{subfigure}
\begin{document}
\begin{flushleft}
{\small\bf BU-HEPP-17-04, Dec. 2017}
\end{flushleft}
\title{IR-improved DGLAP parton shower effects in W~+~jets in pp collisions at $\sqrt{s}=7$ TeV}

\author{B. Shakerin$^{a}$}
\email{bahram\_shakerin@baylor.edu}

\author{B.F.L. Ward$^{a}$}
\email{bfl\_ward@baylor.edu}

\affiliation{$^{a}$Physics Department, Baylor
University, Waco, TX 76798-7316, USA}
\date{\today}

\begin{abstract}
We use HERWIRI1.031, a new Monte Carlo event generator for hadron-hadron scattering at high energies, to study the phenomenological effects of our approach of exact amplitude-based resummation in precision QCD calculations. W~+  jet(s) events with exact NLO QCD corrections are generated in the MG5\_aMC@NLO framework and showered by both HERWIRI1.031 and HERWIG6.5 with PTRMS~=~0 and PTRMS~=~2.2~GeV/c, respectively. Here, PTRMS is the rms value of the intrinsic Gaussian transverse momentum distribution for the partons inside the proton. The differential cross sections for many observables are presented such as the jet rapidities and the jet transverse momenta as well as other event observables such as the scalar sums of transverse momenta of the jets, the missing transverse energy of the jets and the dijets' observables. Finally, we compare our results with the ATLAS and CMS measurements of the W production cross sections in association with jets.  
\end{abstract}
\maketitle
\thispagestyle{plain}
\pagestyle{plain}
\section{Introduction}
In the precision theory of the Standard Model (SM), since we are dealing with the computation of the higher order Feynman diagrams in which the virtual and real radiative corrections are involved, the treatment of the ultraviolet (UV), infrared (IR) and collinear singularities plays a crucial role. The UV singularities appear in the virtual diagrams and are removed by renormalization \cite{tHooft:1,tHooft:2,tHooft:3}. The soft (IR) and collinear singularities appear in theories with massless particles. The IR singularities are removed at the first order of perturbative expansion by Bloch-Nordsieck approach \cite{Bloch:1937pw}. The most general treatment of the IR singularities was developed by Yennie-Frautschi-Suura (YFS) \cite{Yennie:1,Yennie:2}. The main feature of the YFS approach is based on the separation of the infrared divergences as multiplicative exponentiated factors, which are treated exactly to all orders of perturbation theory, and the conversion of the residual exact perturbation expansion into one which has no infrared divergence and, hence, no need for an infrared cutoff. The significant advantage of the YFS formalism is that it is exact to all orders in the QED coupling constant. The YFS formalism was developed and extended by one of us, B.F.L. Ward, to the non-Abelian gauge theories \cite{Ward:1,Ward:2,Ward:3}. One can show that the exact, amplitude-based resummation leads to the IR-improvement of the usual DGLAP-CS theory \cite{Altarelli:1977zs,Dokshitzer,Gribov,Collins} which results in a new set of kernels, parton distributions and attendant reduced cross sections, so that the QCD perturbative results for the respective hadron-hadron or lepton-hadron cross section are unchanged order-by-order in $\alpha_{s}$ at large squared-momentum transfers. This IR-improved behavior, for example, results in kernels that are integrable in the IR limit and therefore are more amenable to realization by the Monte Carlo (MC) method \cite{Ward:4,Ward:5,Ward:6,Ward:7,Ward:8,Ward:9,Ward:10} to arbitrary precision. The advantage of this IR-improved method is better control on the accuracy of a given fixed-order calculation throughout the entire phase space of the respective physical process, especially when the prediction is given by the MC method. This new approach seems important especially in the era of LHC, in which we must deal with the requirements of precision QCD, which involves predictions for QCD processes at the total precision tag of 1\% or better.\par
In this paper, we extend the studies in Refs.~\cite{Ward:4,Ward:5,Ward:6,Ward:7,Ward:8,Ward:9,Ward:10}, which were focused on the single $Z/\gamma^*$ production at FNAL and LHC, to the single W production at the LHC, with the additional change that we look into the properties of
jets, produced in association with the W, in relation to the physics of IR-improved DGLAP-CS kernels. We study whether the 
manifestation of the IR-improved kernels as seen in the decay lepton observables in Refs.~\cite{Ward:4,Ward:5,Ward:6,Ward:7,Ward:8,Ward:9,Ward:10} will also be seen
in the distributions of jet observables. We thus focus on the processes $pp\rightarrow \mathrm{W} + n \mathrm{jets}, n=1,2,3$. We use the MG5\_aMC@NLO~\cite{Alwall:2014hca} framework into which we have introduced the Herwiri1.031~\cite{Ward:4,Ward:5,Ward:6,Ward:7,Ward:8,Ward:9,Ward:10} IR-improved shower to be compared with the standard unimproved Herwig6.5~\cite{Corcella:2000bw} shower in that framework. In this way, we realize exact NLO matrix element matched parton showers with and without IR-improvement. We compare with the data from ATLAS and CMS at 7 TeV to make contact with observations.\par
The paper is organized as follows. In the next section we give a brief review of exact $QED\otimes QCD$ resummation theory. In Section 3 we describe our event generation, analysis and cuts. In Section 4 we compare our predictions with the ATLAS 7 TeV data. In Section 5 we compare our predictions with the CMS 7 TeV data. Section 6 contains our concluding remarks.

\section{Extension of YFS Theory to $QED\otimes QCD$}
We start with a prototypical process $pp\rightarrow W^{\pm}+n(\gamma)+m(g)+X\rightarrow l^{\pm}+\nu_{l^{\pm}}+n'(\gamma)+m(g)+X'$, where $l=\{e,\mu\}$, $\nu_{l^{+}}=\nu_{l}$, and $\nu_{l^{-}}=\bar{\nu}_{l}$. The new $QED\otimes QCD$ YFS extension is obtained by simultaneously resumming the large IR terms in QCD and the IR dominant terms in QED. One can prove that the exponentiated cross section is given by \cite{Ward:11,Ward:12,Ward:13,Ward:14,Ward:15}
\begin{equation}\label{eq1}
    \begin{split}
        d\hat{\sigma}_{exp}&=\sum_{n=0}^{\infty}d\tilde\sigma^{n}=e^{\mathrm{SUM}_{\mathrm{IR}}(\mathit{QCED})}\sum_{n,m=0}^{\infty}\int\prod_{j_{1}=1}^{n}\frac{d^3k_{j1}}{k_{j1}}\prod_{j_{2}=1}^{m}\frac{d^3k^{'}_{j2}}{k^{'}_{j2}}\\
   &\times\int\frac{d^4y}{(2\pi)^4}e^{iy\cdot((p_{1}+q_{1}-p_{2}-q_{2}-\sum k_{j1}-\sum k^{'}_{j2})+D_{\mathit{QCED}}}\\
   &\times \tilde{\bar{\beta}}_{n,m}(k_{1},\ldots,k_{n};k^{'}_{1},\ldots,k^{'}_{m})\frac{d^3p_{2}}{p^{0}_{2}}\frac{d^3q_{2}}{q^{0}_{2}},
    \end{split}
\end{equation}
with $n(\gamma)$ hard photons and $m(g)$ hard gluons, where $\tilde{\bar{\beta}}_{n,m}(k_{1},...,k_{n};k^{'}_{1},...,k^{'}_{m})$ are the YFS residuals which are free of all infrared divergences to all orders in $\alpha_{s}$ and $\alpha$. The infrared functions are given by
\begin{align}
   \mathrm{SUM}_{\mathrm{IR}}(\mathit{QCED})&=2\alpha_{s}ReB^{nls}_{\mathit{QCED}}+2\alpha_{s}\tilde{B}^{nls}_{\mathit{QCED}}(K_{max}),\\
2\alpha_{s}\tilde{B}_{\mathit{QCED}}(K_{max})&=\int\frac{d^3k}{k^0}\tilde{S}^{nls}_{\mathit{QCED}}(k)\theta(K_{max}-k),\\
D_{\mathit{QCED}}=\int\frac{d^3k}{k}&\tilde{S}^{nls}_{\mathit{QCED}}(k)\bigg[e^{-iy\cdot k}-\theta(K_{max}-k)\bigg], 
\end{align}
and the functions $\mathrm{SUM}_{\mathrm{IR}}(\mathit{QCED})$, $D_{\mathit{QCED}}$ are determined form their QCD analougs $\mathrm{SUM}_{\mathrm{IR}}(\mathit{QCD})$, $D_{\mathit{QCD}}$ via the following substitutions
\begin{equation}\label{eq5}
 \left\{
 \begin{array}{l}
    B^{nls}_{\mathit{QCD}}\rightarrow B^{nls}_{\mathit{QCD}}+B^{nls}_{\mathit{QED}}\equiv B^{nls}_{\mathit{QCED}}, \\ 
    \tilde{B}^{nls}_{\mathit{QCD}}\rightarrow \tilde{B}^{nls}_{\mathit{QCD}}+\tilde{B}^{nls}_{\mathit{QED}}\equiv \tilde{B}^{nls}_{\mathit{QCED}},\\
    \tilde{S}^{nls}_{\mathit{QCD}}\rightarrow \tilde{S}^{nls}_{\mathit{QCD}}+\tilde{S}^{nls}_{\mathit{QED}}\equiv \tilde{S}^{nls}_{\mathit{QCED}}.
 \end{array}
 \right.
\end{equation}
In Eq (\ref{eq5}), the superscript $nls$ asserts that the infrared functions $B_{\mathit{QCD}}$, $B_{\mathit{QED}}$, $\tilde{B}_{\mathit{QCD}}$, $\tilde{B}_{\mathit{QED}}$ and $\tilde{S}_{\mathit{QCD}}$ are DGLAP-CS synthesized. These infrared functions have been introduced in Ref. \cite{Ward:16,Ward:17,Ward:18,Ward:19}. The QCD exponentiation of the master formula in Eq (\ref{eq1}) leads to a new set of IR-improved splitting functions listed below
\begin{equation}\label{eq6}
    \left\{ 
    \begin{array}{l}
        P^{exp}_{qq}(z)=C_{F}~e^{\frac{1}{2}\delta_{q}}F_{\mathit{YFS}}(\gamma_{q})\bigg[\frac{1+z^2}{1-z}(1-z)^{\gamma_{q}}-f_{q}(\gamma_{q})\delta(1-z)\bigg],\\
        P^{exp}_{Gq}(z)=C_{F}~e^{\frac{1}{2}\delta_{q}}F_{\mathit{YFS}}(\gamma_{q})\frac{1+(1-z)^2}{z}~z^{\gamma_{q}},\\
         P^{exp}_{qG}(z)=e^{\frac{1}{2}\delta_{q}}F_{\mathit{YFS}}(\gamma_{q})\frac{1}{2}\bigg\{z^2(1-z)^{\gamma_{G}}+(1-z)^2z^{\gamma_{G}}\bigg\},\\
        \begin{split}
        P^{exp}_{GG}(z)&=2C_{G}F_{\mathit{YFS}}(\gamma_{G})e^{\frac{1}{2}\delta_{G}}\bigg\{\frac{1-z}{z}z^{\gamma_{G}}+\frac{z}{1-z}(1-z)^{\gamma_{G}}\\+&\frac{1}{2}\bigg((1-z)z^{\gamma_{G}+1}+z(1-z)^{\gamma_{G}+1}\bigg)-f_{G}(\gamma_{G})\delta(1-z)\bigg\},\\
    \end{split}
    \end{array} 
        \right.
\end{equation}
where
\begin{equation}
\left\{ 
    \begin{array}{l}
    \begin{split}
       \gamma_{q}&=C_{F}\frac{\alpha_{s}}{\pi}t=\frac{4C_{F}}{\beta_{0}}~,~\delta_{q}=\frac{\gamma_{q}}{2}+\frac{\alpha_{s}C_{F}}{\pi}\bigg(\frac{\pi^2}{3}-\frac{1}{2}\bigg),\\
       \gamma_{G}&=C_{G}\frac{\alpha_{s}}{\pi}t=\frac{4C_{G}}{\beta_{0}}~,~\delta_{G}=\frac{\gamma_{G}}{2}+\frac{\alpha_{s}C_{G}}{\pi}\bigg(\frac{\pi^2}{3}-\frac{1}{2}\bigg),\\
       F_{YFS}(x)&=\frac{e^{C_{E}x}}{\Gamma(1+x)}~,~\beta_{0}=11-\frac{2}{3}n_{f}=4\beta_{1}~,~C_{E}=0.57721566\ldots,\\
       f_{q}(\gamma_{q})&=\frac{2}{\gamma_{q}}-\frac{2}{\gamma_{q}+1}+\frac{1}{\gamma_{q}+2},\\
       \bar{f}_{G}(\gamma_{G})&=\frac{n_{f}}{C_{G}}\frac{1}{(1+\gamma_{G})(2+\gamma_{G})(3+\gamma_{G})}+\frac{2}{\gamma_{G}(1+\gamma_{G})(2+\gamma_{G})}\\
      &+\frac{1}{(1+\gamma_{G})(2+\gamma_{G})}+\frac{1}{2(3+\gamma_{G})(4+\gamma_{G})}+\frac{1}{(2+\gamma_{G})(3+\gamma_{G})(4+\gamma_{G})}.
    \end{split}
     \end{array} 
        \right.
\end{equation}
\vspace{2ex}
Finally, for precision LHC theory, the famous factorization theorem \cite{Ellis:1978ty}
\begin{equation}
\sigma=\sum_{i,j}{}\int dx_{1}~dx_{2}F_{i}(x_{1})F_{j}(x_{2})\hat{\sigma}(x_{1}x_{2}s),
\end{equation}
is written in the following form
\begin{equation}
\sigma=\sum_{i,j}{}\int dx_{1}~dx_{2}F^{'}_{i}(x_{1})F^{'}_{j}(x_{2})\hat{\sigma}^{'}(x_{1}x_{2}s)
\end{equation}
where the primed quantities are associated with the kernels and cross sections derived in Eqs~(\ref{eq6}) and (\ref{eq1}) respectively. 
The implementation of the new IR-improved kernels in the HERWIG6.5 \cite{Corcella:2000bw} environment leads to a new MC, HERWIRI1.031, as described in Ref.~\cite{Joseph:2010cq}.
In what follows, we present results using both the original Herwig6.5 and the new IR-improved Herwiri1.031. For both MG5\_aMC@NLO/HERWIG and MG5\_aMC@NLO/HERWIRI simulations, we use the NNPDF2.3nlo PDF's \cite{Ball:2012cx}. \par
\section{Event Generation, Analysis and Cuts}
The generators for W~+ jet events are MADGRAPH5\_aMC@NLO \cite{Alwall:2014hca} interfaced with HERWIG6.521 and HERWIRI1.031, which use with exact next-to-leading-order (NLO) matrix element calculations matched to the respective parton shower. The number of events generated for the W, W~+~1 jet, W~+~2 jets, and W~+~3 jets processes are $10^7$, $10^6$, $10^5$, and $10^5$, respectively.\,These events are showered by MADGRAPH5\_aMC@NLO/HERWIRI1.031\footnote{Note that only the showers are IR-improved in Herwiri1.031 and that, since this affects terms starting at $O(\alpha_s^2L)$, exactness at $O(\alpha_{s})$ is unaffected in MADGRAPH5\_aMC@NLO/Herwiri1.031.} (PTRMS = 0) and MADGRAPH5\_aMC@NLO/HERWIG6.521 (PTRMS = 2.2 GeV).\footnote{We will see later that HERWIRI gives either a better fit to the data or an acceptable fit without this extra intrinsic Gaussian kick.} During the analysis, jets were reconstructed using the anti-$k_{t}$ algorithm with FastJet \cite{Cacciari:2011ma} and the cuts in Tables~\ref{t1} and~\ref{t2} were imposed for the ATLAS and CMS results, respectively. 

\begin{table}[h!]
\centering 
\begin{tabular}{ p{6cm}p{6cm}  }
\hline
\multicolumn{2}{c}{Combined channel~~~$W\rightarrow l+\nu_{l}$~~where~$l=\{e,\mu\}$}\\
\hline
Lepton $P^{l}_{T}$ & $P^{l}_{T}>25~\mathrm{GeV}$  \\

Lepton rapidity $\eta_{l}$ & $|\eta_{l}|<2.5$  \\

Missing transverse energy  &$E^{\mathrm{miss}}_{T}>25~\mathrm{GeV}$  \\

Transverse mass    &$m_{T}>40~\mathrm{GeV}$ \\

Jet algorithm & Anti-$k_{T}$\\

Radius parameter~$R$& $R=0.4$\\

Jet $P^{\mathrm{jet}}_{T}$ & $P^{\mathrm{jet}}_{T}>30~\mathrm{GeV}$ \\

Jet rapidity $Y_{\mathrm{jet}}$ & $|Y_{\mathrm{jet}}|<4.4$  \\

Jet isolation& $\Delta R(l,\mathrm{jet})>0.5$~(jet is removed)\\
\hline
\end{tabular}
\caption{Kinematic criteria defining the fiducial phase space for the $W\rightarrow l+\nu_{l}$ channel.}
\label{t1}
\end{table}
\begin{table}[h!]
\centering 
\begin{tabular}{ p{6cm}p{6cm} }
\hline
\multicolumn{2}{c}{Muon channel~~~($W\rightarrow \mu+\nu_{\mu}$)}\\
\hline
Lepton $P^{\mu}_{T}$ & $P^{\mu}_{T}>25~\mathrm{GeV}$  \\

Lepton rapidity $\eta_{\mu}$ & $|\eta_{\mu}|<2.1$  \\

Missing transverse energy  &$E^{\mathrm{miss}}_{T}>25~\mathrm{GeV}$  \\

Transverse mass    &$m_{T}>50~\mathrm{GeV}$ \\

Jet algorithm & Anti-$k_{t}$\\

Radius parameter~$R$& $R=0.5$\\

Jet $P^{jet}_{T}$ & $P^{\mathrm{jet}}_{T}>30~\mathrm{GeV}$ \\

Jet pseudorapidity $\eta_{\mathrm{jet}}$ & $|\eta_{\mathrm{jet}}|<2.4$  \\

Jet isolation& $\Delta R(\mu,\mathrm{jet})>0.5$~(jet is removed)\\
\hline
\end{tabular}
\caption{Kinematic criteria defining the fiducial phase space for the $W\rightarrow \mu+\nu_{\mu}$~~channel }
\label{t2}
\end{table}

The transverse mass, $m_{T}$, is defined as $m_{T}=\sqrt{2P^{l}_{T}P^{\nu_{l}}_{T}(1-\cos\Delta\phi})$ where $\Delta\phi$ is the difference in the azimuthal angle between the direction of the lepton momentum and the associated neutrino, $\nu_{l}$, which can be written as
\begin{equation}
\Delta\phi=\phi^{l}-\phi^{\nu_{l}}.
\end{equation}
Rapidity is defined as $\displaystyle\frac{1}{2}\ln\left[\frac{E+p_{z}}{E-p_{z}}\right]$, where $E$ denotes the energy of the particle and $p_{z}$ is the longitudinal component of the momentum. Finally, the jet isolation, $\Delta R$, which is a Lorentz invariant quantity for massless particles, is defined as 
\begin{equation}
\Delta R(l,\mathrm{jet})=\sqrt{\Delta\phi^2(l,\mathrm{jet})+\Delta\eta^2(l,\mathrm{jet})},
\end{equation}
where\vspace{2mm}
\begin{equation}
\left\{ \begin{array}{ll}
         \Delta\phi(l,\mathrm{jet})=\phi_{l}-\phi_\mathrm{jet},\\
         \Delta\eta(l,\mathrm{jet})=\eta_{l}-\eta_\mathrm{jet},\\
         \eta=-\ln\tan\left(\frac{\theta}{2}\right),\end{array} \right.
\label{etadef}
\end{equation}
where $\theta$ is the angle between the respective particle three-momentum $\vec{P}$ and the positive direction of the beam axis. The $E^{\mathrm{miss}}_{T}$ is calculated as the negative vector sum of the transverse momenta of calibrated leptons, photons and jets and additional low-energy deposits in the calorimeter. 

\section{Results (ATLAS Collaboration)}
In this section, the measured W$(\rightarrow l+\nu_{l})$~+ jets fiducial cross sections~\cite{Aad:2014qxa} are shown and compared to the predictions of MADGRAPH5\_aMC@NLO/HERWIRI1.031 and MADGRAPH5\_aMC@NLO/HERWIG6.521. Each distribution is combined separately by minimizing a $\chi^2$ function. The factors applied to the theory predictions are summarized in Appendix~A and Appendix~B.  \\ \vspace{2mm}
We have used the following notation throughout this paper:
\begin{itemize}
 \item herwiri~$\equiv$~MADGRAPH5\_aMC@NLO/HERWIRI1.031 (PTRMS~=~0);
 \item herwig~ $\equiv$~ MADGRAPH5\_aMC@NLO/HERWIG6.521 (PTRMS~=~2.2~GeV).
\end{itemize}

\subsection{Transverse Momentum Distributions}
\begin{figure}[h]
\centering
\includegraphics[scale=0.4]{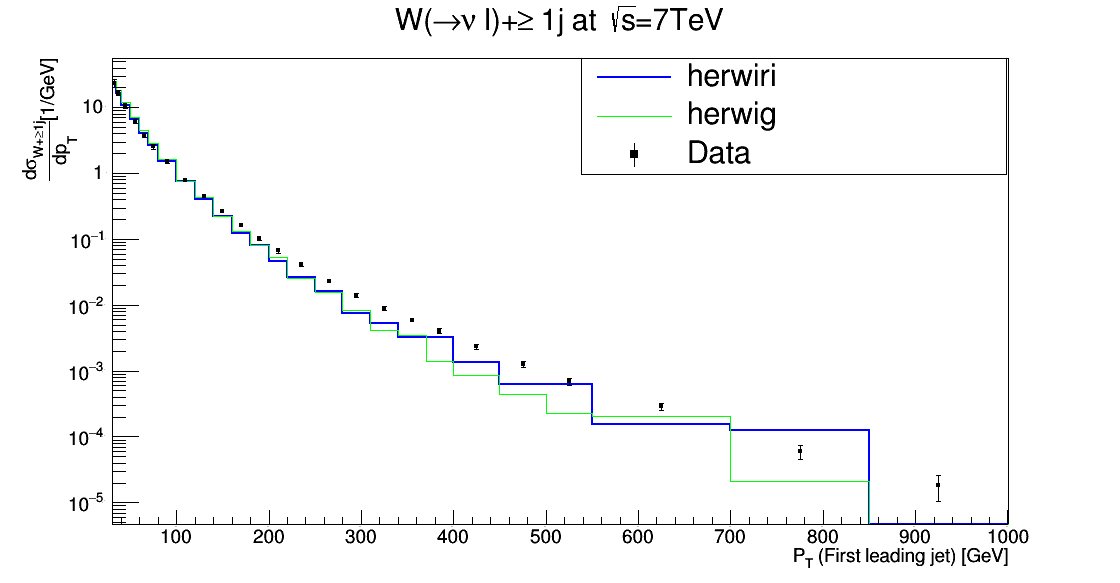}
\caption{Cross section for the production of W~+ jets as a function of the leading-jet $P_{T}$ in $N_{jet}\geq 1$. The data are compared to predictions from MADGRAPH5\_aMC@NLO/HERWIRI1.031 and MADGRAPH5\_aMC@NLO/HERWIG6.521.}
\label{fig1}
\end{figure}

\begin{figure}[H]
\includegraphics[scale=0.4]{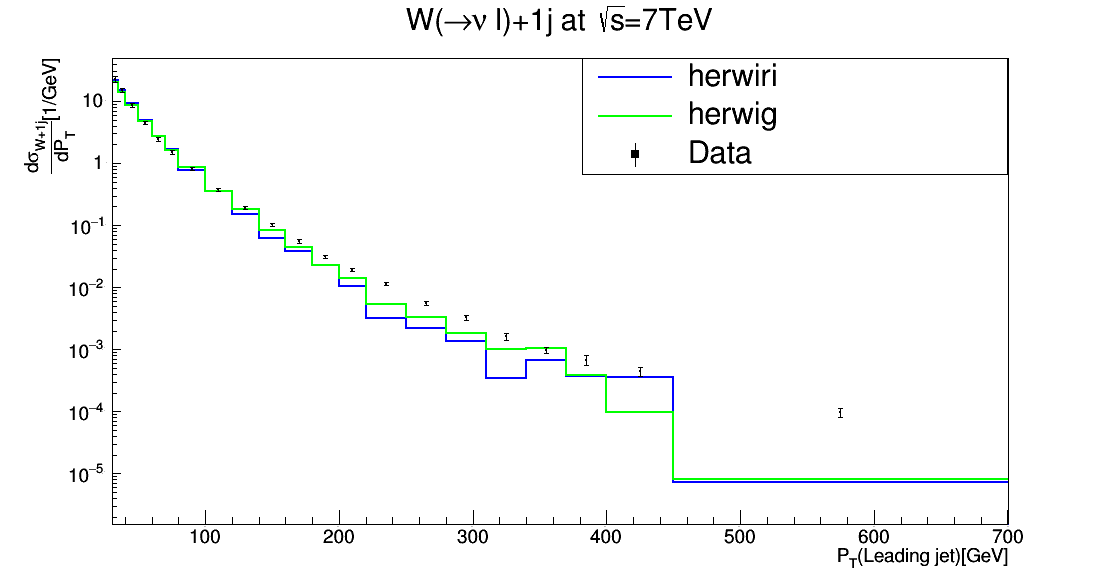}
\caption{Cross section for the production of W~+ jets as a function of the leading-jet $P_{T}$ in $N_{jet}=1$. The data are compared to predictions from MADGRAPH5\_aMC@NLO/HERWIRI1.031 and MADGRAPH5\_aMC@NLO/HERWIG6.521.}
\label{fig2}
\end{figure}

The differential cross sections as a function of the leading jet transverse momentum are shown in Figure.~\ref{fig1} and Figure.~\ref{fig2} for the W~+~$\geq$1 jet and W~+ 1 jet cases, respectively. In both cases, there is agreement between the data and predictions provided by HERWIRI and HERWIG in the soft regime.
\par
In Figure.~\ref{fig1}, for $P_{T}< 140~\mathrm{GeV}$, HERWIRI predictions are in better agreement with the data, where $\big(\frac{\chi^2}{d.o.f}\big)_{\texttt{HERWIRI}}=0.76$ and $\big(\frac{\chi^2}{d.o.f}\big)_{\texttt{HERWIG}}=2.04$. The $\big(\frac{\chi^2}{d.o.f}\big)$ functions have been calculated for the first 9 bins. In Figure.~\ref{fig2}, for $P_{T}< 120~\mathrm{GeV}$, $\big(\frac{\chi^2}{d.o.f}\big)_{\texttt{HERWIRI}}=1.13$ and $\big(\frac{\chi^2}{d.o.f}\big)_{\texttt{HERWIG}}=0.96$. The $\big(\frac{\chi^2}{d.o.f}\big)$ functions have been calculated for the first 8 bins. 
\begin{figure}[H]
\centering
\includegraphics[scale=0.4]{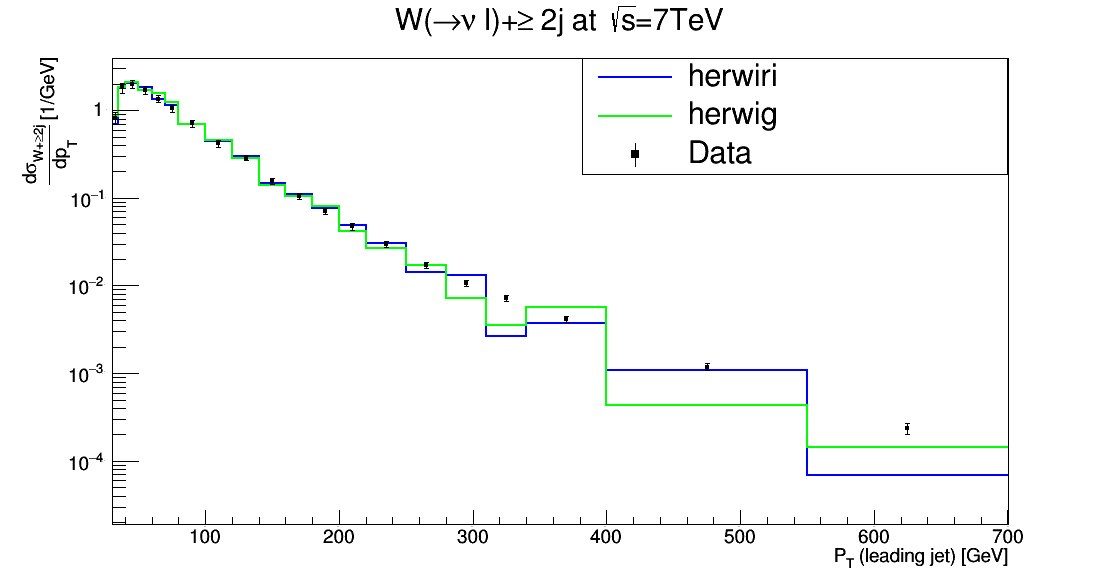}
\caption{Cross section for the production of W~+ jets as a function of the leading-jet $P_{T}$ in $N_{jet}\geq 2.$ The data are compared to predictions from MADGRAPH5\_aMC@NLO/HERWIRI1.031 and MADGRAPH5\_aMC@NLO/HERWIG6.521.}
\label{fig3}
\end{figure}
\begin{figure}[H]
\includegraphics[scale=0.4]{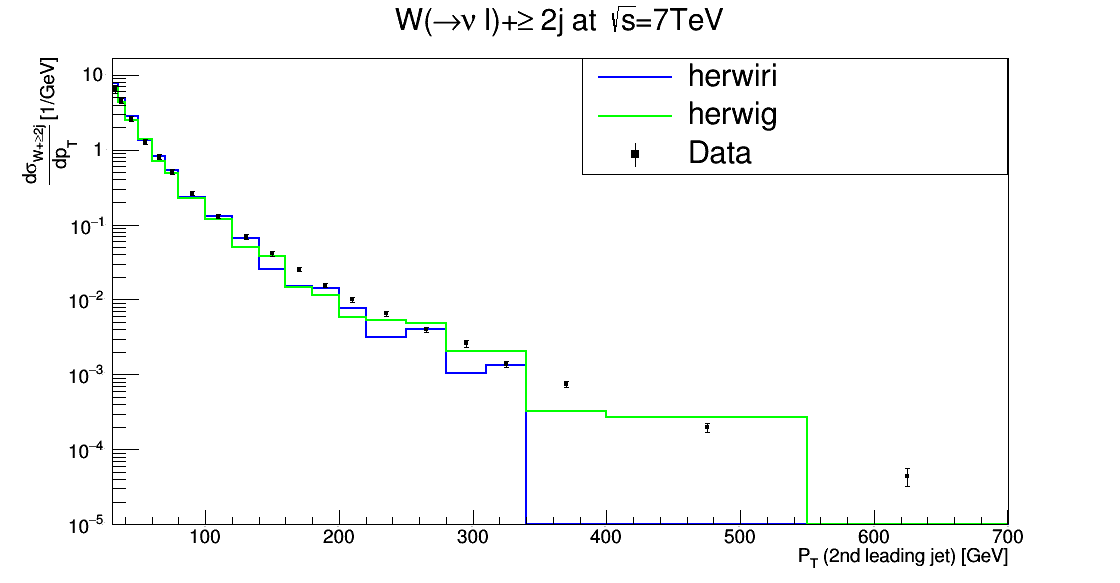}
\caption{Cross section for the production of W~+ jets as a function of the second leading-jet $P_{T}$ in $N_{jet}\geq 2.$ The data are compared to predictions from MADGRAPH5\_aMC@NLO/HERWIRI1.031 and MADGRAPH5\_aMC@NLO/HERWIG6.521.}
\label{fig4}
\end{figure}
For the sake of clarification, the ratio plots for Figure. 1 and Figure. 2 are given in Appendix C. In the ratio plot, each point represents $\mathrm{\frac{Data}{Theory}}$ (See Figures 41 to 48).

The differential cross sections for the production of W~+~$\geq$2 jets as a function of the leading jet $P_{T}$ and the second leading jet $P_{T}$ are shown in Figure.~\ref{fig3} and Figure.~\ref{fig4}, respectively. HERWIRI and HERWIG generally describe the data well for $P_{T}<200~\mathrm{GeV}$. In Figure.~\ref{fig3}, $\big(\frac{\chi^2}{d.o.f}\big)_{\texttt{HERWIRI}}=1.19$ and $\big(\frac{\chi^2}{d.o.f}\big)_{\texttt{HERWIG}}=1.49$, while for $200<P_{T}<350~\mathrm{GeV}$ it seems that they both fail to describe the data. For $250<P_{T}<550~\mathrm{GeV}$, HERWIRI predictions overlap with the data while HERWIG either underestimates or overestimates the data. Finally, for energies higher than $550~\mathrm{GeV}$, they both underestimate the data. The behaviors for $P_{T} > 200~\mathrm{GeV}$ are consistent with our theoretical curves' exact NLO Matrix Element (ME) matched parton shower precision.
\par
Figure.~\ref{fig4} shows that HERWIRI, in general, gives a better fit to the data for $P_{T}<150~\mathrm{GeV}$, where $\big(\frac{\chi^2}{d.o.f}\big)_{\texttt{HERWIRI}}=1.06$ and $\big(\frac{\chi^2}{d.o.f}\big)_{\texttt{HERWIG}}=1.69$. For higher $P_{T}$, in some cases HERWIRI predictions overlap with the data while HERWIG either underestimates or overestimates the data. We conclude that HERWIRI gives a better fit to the data in the soft regime as expected.
\begin{figure}[b]
\centering
\includegraphics[scale=0.4]{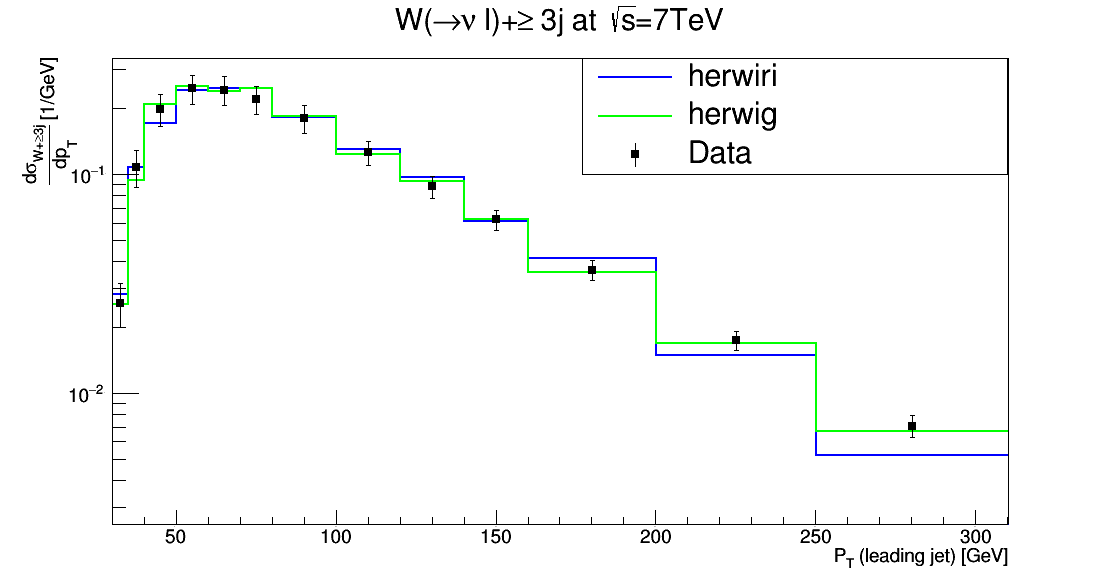}
\caption{Cross section for the production of W~+ jets as a function of the leading-jet $P_{T}$ in $N_{jet}\geq 3.$ The data are compared to predictions from MADGRAPH5\_aMC@NLO/HERWIRI1.031 and MADGRAPH5\_aMC@NLO/HERWIG6.521.}
\label{fig5}
\end{figure}
\begin{figure}[H]
\includegraphics[scale=0.4]{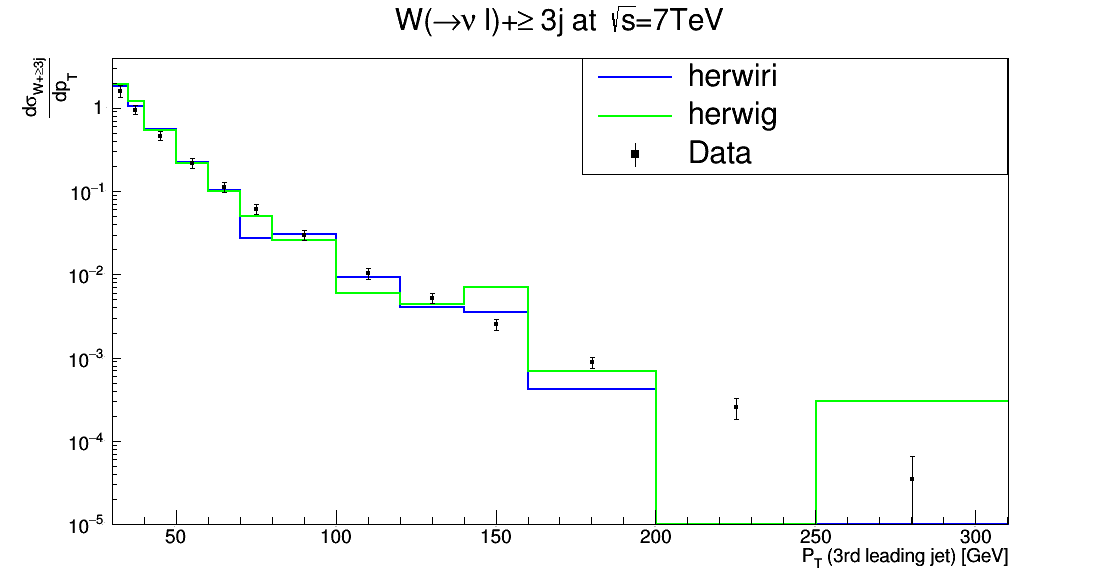}
\caption{Cross section for the production of W~+ jets as a function of the third leading-jet $P_{T}$ in $N_{jet}\geq 3.$ The data are compared to predictions from MADGRAPH5\_aMC@NLO/HERWIRI1.031 and MADGRAPH5\_aMC@NLO/HERWIG6.521.}
\label{fig6}
\end{figure}
The differential cross sections for the production of W~+~$\geq$3 jets as a function of the leading jet $P_{T}$ and the third leading jet $P_{T}$ are shown in Figure.~\ref{fig5} and Figure.~\ref{fig6}, respectively. In Figure.~\ref{fig5}, for $P_{T}<150~\mathrm{GeV}$, the predictions provided by HERWIRI and HERWIG are in complete agreement with the data, where $\big(\frac{\chi^2}{d.o.f}\big)_{\texttt{HERWIRI}}=0.27$ and $\big(\frac{\chi^2}{d.o.f}\big)_{\texttt{HERWIG}}=0.20$. For $P_{T}>150~\mathrm{GeV}$, HERWIG gives a better fit to the data while HERWIRI underestimates the data. In Figure.~\ref{fig6}, HERWIRI gives a better fit to the data for low $P_{T}$, $P_{T}<150~\mathrm{GeV}$, where $\big(\frac{\chi^2}{d.o.f}\big)_{\texttt{HERWIRI}}=3.27$ and $\big(\frac{\chi^2}{d.o.f}\big)_{\texttt{HERWIG}}=3.97$. For large $P_{T}$, in almost all cases HERWIRI and HERWIG predictions either underestimate or overestimate the data.\par
In general, one could conclude that the predictions provided by HERWIRI give as good a fit or a better fit to the data for 
soft $P_{T}$ without the need of an 'ad hoc' intrinsic Gaussian rms transverse momentum of 2.2 GeV as needed by HERWIG.
\subsection{Rapidity Distributions}
\begin{figure}[H]
\centering
\includegraphics[scale=0.4]{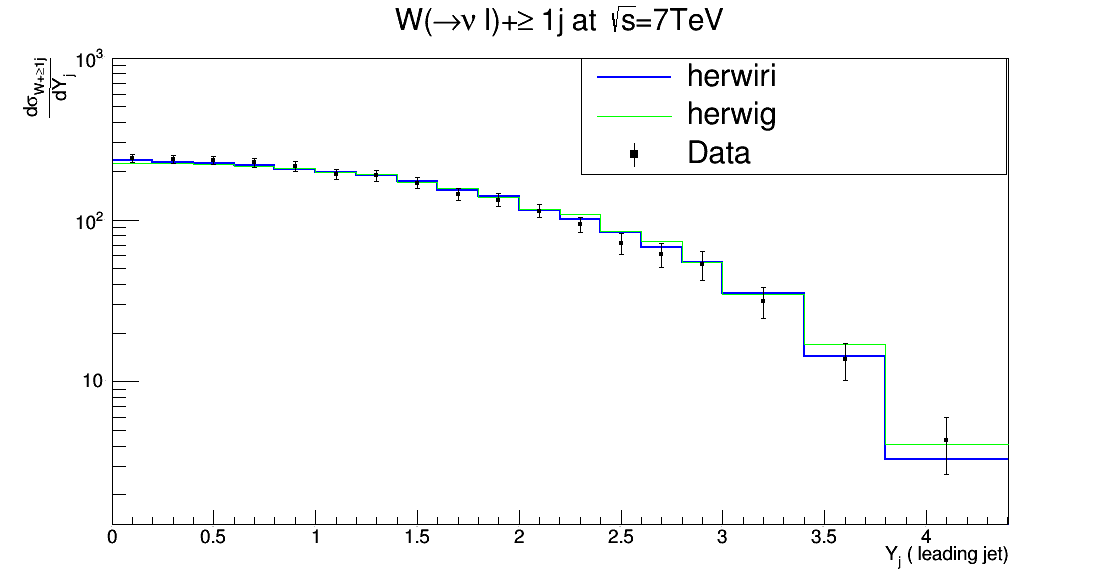}
\caption{Cross section for the production of W~+ jets as a function of the leading-jet $Y_{j}$ in $N_{jet}\geq 1.$ The data are compared to predictions from MADGRAPH5\_aMC@NLO/HERWIRI1.031 and MADGRAPH5\_aMC@NLO/HERWIG6.521.}
\label{fig7}
\end{figure}

The differential cross sections for the production of W~+~$\geq$1 jet as a function of the leading jet $Y_{j}$ are shown in Figure.~\ref{fig7}. The predictions provided by HERWIRI and HERWIG are generally in agreement with the data, although in three cases HERWIRI predictions overlap with the data while the HERWIG predictions either underestimate or overestimate the data. We clearly conclude that HERWIRI and HERWIG give a very good fit to the data with $\big(\frac{\chi^2}{d.o.f}\big)_{\texttt{HERWIRI}}=0.35$ and $\big(\frac{\chi^2}{d.o.f}\big)_{\texttt{HERWIG}}=0.70$.\par
\begin{figure}[H]
\includegraphics[scale=0.4]{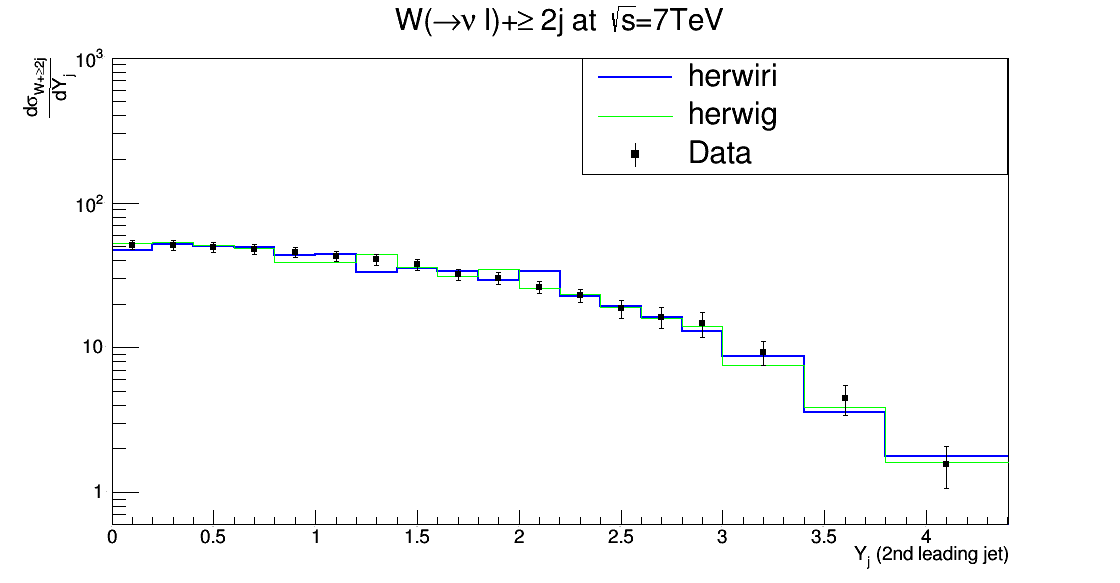}
\caption{Cross section for the production of W~+ jets as a function of the second leading-jet $Y_{j}$ in $N_{jet}\geq 2.$ The data are compared to predictions from MADGRAPH5\_aMC@NLO/HERWIRI1.031 and MADGRAPH5\_aMC@NLO/HERWIG6.521.}
\label{fig8}
\end{figure}

 
The differential cross sections for the production of W~+ $\geq$2 jets as a function of the second leading jet $Y_{j}$ are shown in Figure.~\ref{fig8}. The results provided by HERWIRI and HERWIG overlap with the data in almost all cases. In two cases, the HERWIRI predictions overlap with the data and in two cases the HERWIG results overlap with the data while HERWIRI predictions either underestimate or overestimate the data: $\big(\frac{\chi^2}{d.o.f}\big)_{\texttt{HERWIRI}}=1.01$ and $\big(\frac{\chi^2}{d.o.f}\big)_{\texttt{HERWIG}}=0.63$. Here, both theoretical predictions give acceptable fits to the data.\par

\begin{figure}[H]
\centering
\includegraphics[scale=0.4]{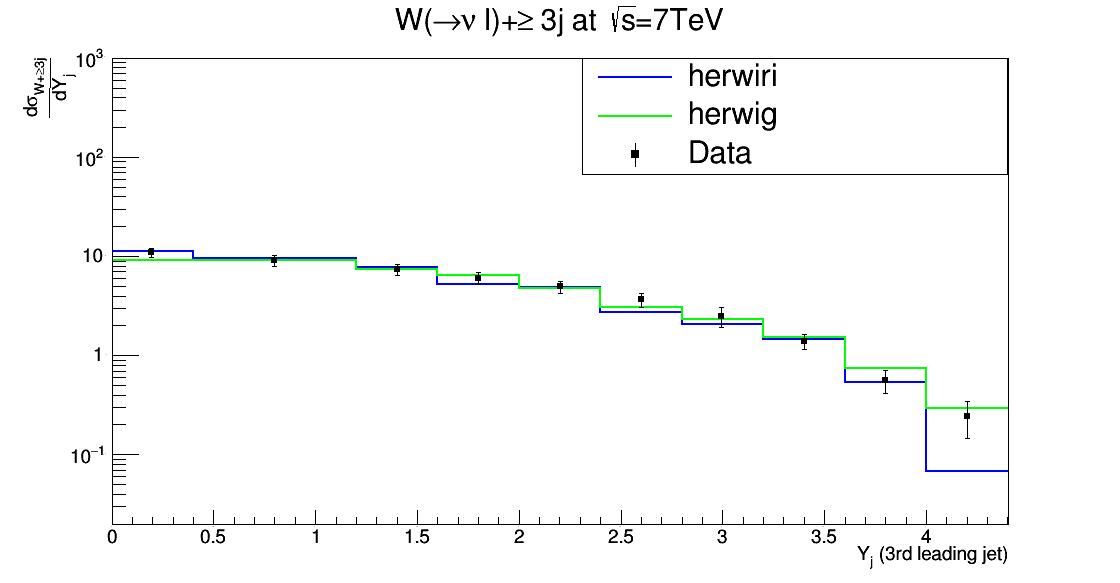}
\caption{Cross section for the production of W~+ jets as a function of the third leading-jet $Y_{j}$ in $N_{jet}\geq 3.$ The data are compared to predictions from MADGRAPH5\_aMC@NLO/HERWIRI1.031 and MADGRAPH5\_aMC@NLO/HERWIG6.521.}
\label{fig9}
\end{figure}
The differential cross sections for the production of W~+~$\geq3$ jets as a function of the third leading jet $Y_{j}$ are shown in Figure.~\ref{fig9}. For $Y_{j}<3.6$, with the exception of one case in which only the HERWIG prediction overlaps with the error bars on the data, HERWIRI and HERWIG predictions are in agreement with the data. For $Y_{j}>3.6$, in one case HERWIRI overlaps with the error bars on the data while HERWIG overestimates the data, and in the other case HERWIG overlaps with the error bars on the data while HERWIRI underestimates the data: $\big(\frac{\chi^2}{d.o.f}\big)_{\texttt{HERWIRI}}=1.05$ and $\big(\frac{\chi^2}{d.o.f}\big)_{\texttt{HERWIG}}=0.43$ so that both predictions give acceptable fits to the data. 


\subsection{Dijet Angular Variables, Invariant Mass, Separation}
In this subsection the differential cross sections are shown as functions of the difference in azimuthal angle ($\Delta\phi_{j_{1},j_{2}}$), the difference in the rapidity ($\Delta Y_{j_{1},j_{2}}$), the angular separation ($\Delta R_{j_{1},j_{2}})$ and the dijet invariant mass ($m_{j_{1},j_{2}}$) in comparison to the data. We define the aforementioned variables as follows
\begin{align}
\Delta Y_{j_{1},j_{2}}&=|Y_{j_{1}}-Y_{j_{2}}|,\\
\Delta\phi_{j_{1},j_{2}}&=|\phi_{j_{1}}-\phi_{j_{2}}|,\\
\Delta R_{j_{1},j_{2}}&=\sqrt{(\Delta\phi_{j_{1},j_{2}})^2+\Delta\eta_{j_{1},j_{2}})^2},\\
M_{j_{1},j_{2}}&=\sqrt{(E_{j_{1}}+E_{j_{2}})^2-(\vec{P}_{j_{1}}+\vec{P}_{j_{2}})^2}=\sqrt{m^2_{j_{1}}+m^2_{j_{2}}+2(E_{j_{1}}E_{j_{2}}-\vec{P}_{j_{1}}\cdot\vec{P}_{j_{2}})}.
\end{align}
We note that in Eq~(15), $\Delta\eta_{j_{1},j_{2}}$ is the difference in pseudorapidity\footnote{The rapidity term in $\Delta R=\sqrt{\Delta\phi^2+\Delta Y^2}$ is often replaced by pseudorapidity if the involved particles are massless} of the first and second leading jets. The $i$th jet is defined as
\begin{equation}
    P^{\mu}_\textit{ith-jet}=(E_{j_{1}}, \vec{P}_\textit{ith-jet})
\end{equation}

\begin{figure}[H]
\centering
\includegraphics[scale=0.4]{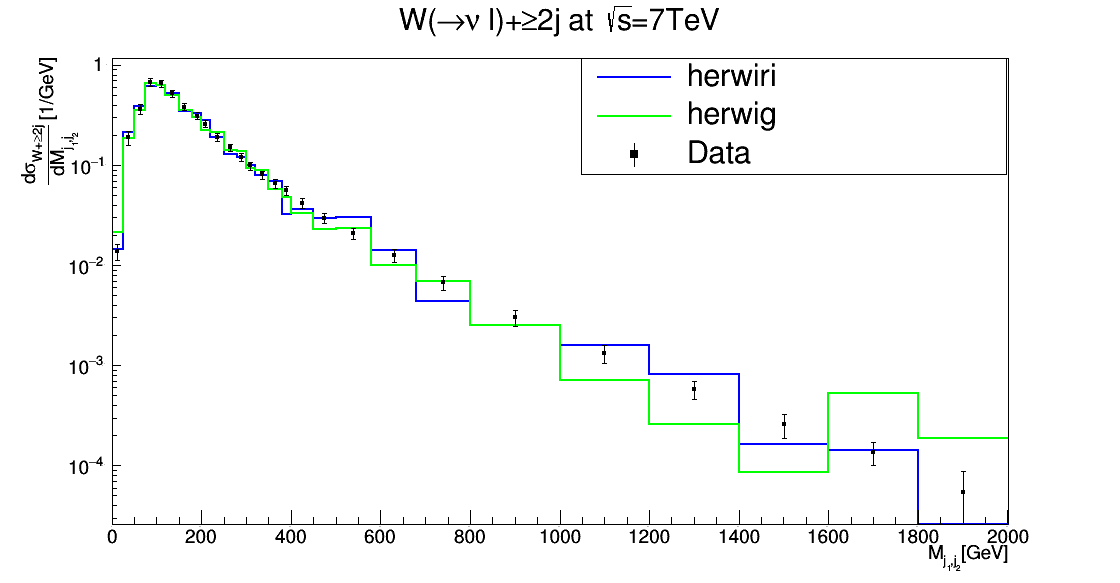}
\caption{Cross section for the production of W~+ jets as a function of the dijet invariant mass $m_{j_{1},j_{2}}$ between the two leading jets in $N_{jet}\geq 2.$ The data are compared to predictions from MADGRAPH5\_aMC@NLO/HERWIRI1.031 and MADGRAPH5\_aMC@NLO/HERWIG6.521.}
\label{fig10}
\end{figure}

The differential cross sections for the production of W~+ $\geq$2 jets as a function of the dijet invariant mass between the two leading jets are shown in Figure.~\ref{fig10}. The cross sections are fairly well modeled by HERWIRI for $M_{j_{1},j_{2}}<300~\mathrm{GeV}$. For $M_{j_{1},j_{2}}>300~\mathrm{GeV}$ there are cases in which HERWIRI gives a good fit to the data while HERWIG predictions either underestimate or overestimate the data. In comparison, predictions provided by HERWIRI describe the data somewhat better than do those provided by HERWIG: $\big(\frac{\chi^2}{d.o.f}\big)_{\texttt{HERWIRI}}=1.18$ and $\big(\frac{\chi^2}{d.o.f}\big)_{\texttt{HERWIG}}=1.69$ for $M_{j_{1},j_{2}}<300~\mathrm{GeV}$ . \par
\begin{figure}[H]
\includegraphics[scale=0.4]{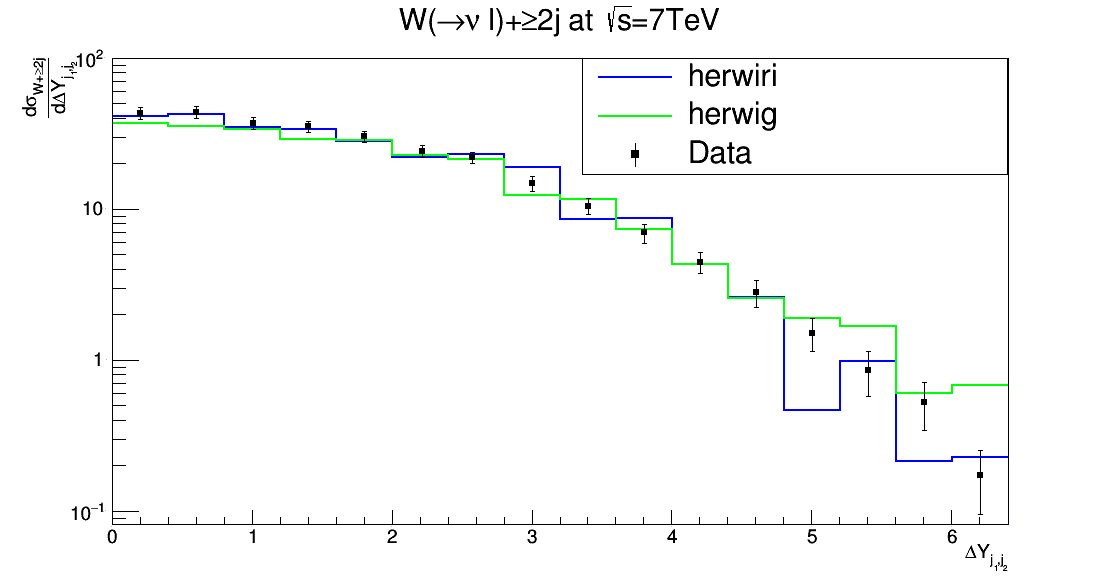}
\caption{Cross section for the production of W~+ jets as a function of the difference in the rapidity between the two leading jets in $N_{jet}\geq 2.$ The data are compared to predictions from MADGRAPH5\_aMC@NLO/HERWIRI1.031 and MADGRAPH5\_aMC@NLO/HERWIG6.521.}
\label{fig11}
\end{figure}

The differential cross sections for the production of W~+~$\geq$2 jets as a function of the difference in the rapidity between the two leading jets are shown in Figure.~\ref{fig11}. For $\Delta Y_{j_{1}j_{2}}<3$ the predictions provided by HERWIRI give a better fit to the data. For $3<\Delta Y_{j_{1}j_{2}}<4$, HERWIG results provide a better description of the data: $\big(\frac{\chi^2}{d.o.f}\big)_{\texttt{HERWIRI}}=2.08$ and $\big(\frac{\chi^2}{d.o.f}\big)_{\texttt{HERWIG}}=4.77$, so that overall HERWIRI gives a better fit to the data.\\
\par
\begin{figure}[h!]
\centering
\includegraphics[scale=0.4]{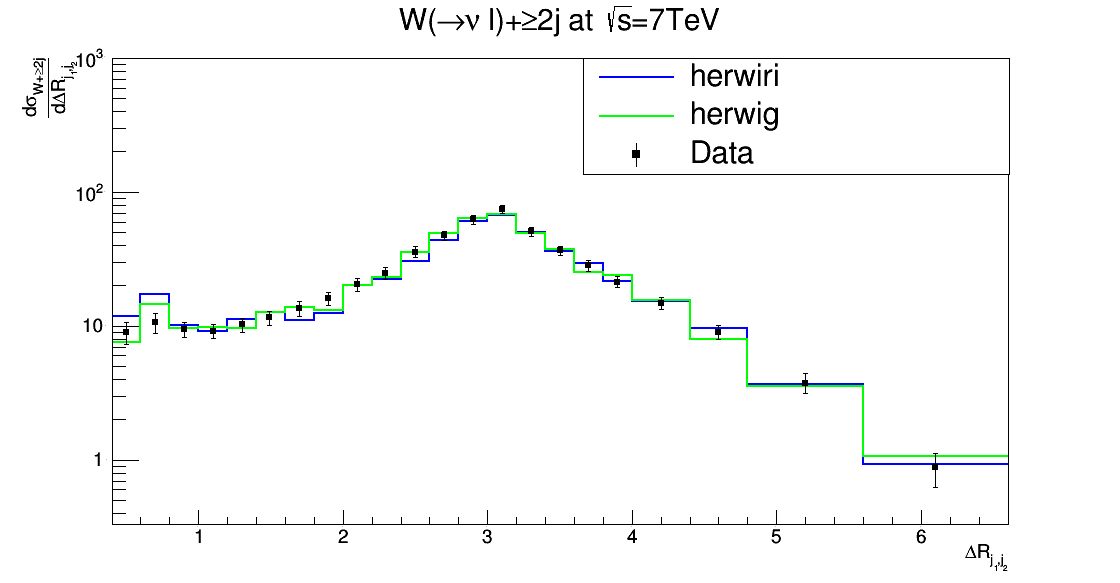}
\caption{Cross section for the production of W~+ jets as a function of the angular separation between the two leading jets for $N_{jet}\geq 2.$ The data are compared to predictions from MADGRAPH5\_aMC@NLO/HERWIRI1.031 and MADGRAPH5\_aMC@NLO/HERWIG6.521.}
\label{fig12}
\end{figure}

The differential cross sections for the production of W~+~$\geq$2 jets as a function of the angular separation between the two leading jets are shown in Figure.~\ref{fig12}. For $\Delta R_{j_{1},j_{2}}>3$, the cross sections are fairly well modeled by the predictions of HERWIRI and HERWIG. For $\Delta R_{j_{1},j_{2}}<3$, in at least two cases the prediction provided by either of them are outside of the error bars on the data; in most cases they both give a satisfactory prediction relative to the data: $\big(\frac{\chi^2}{d.o.f}\big)_{\texttt{HERWIRI}}=1.59$ and $\big(\frac{\chi^2}{d.o.f}\big)_{\texttt{HERWIG}}=0.78$. \par
\begin{figure}[H]
\includegraphics[scale=0.4]{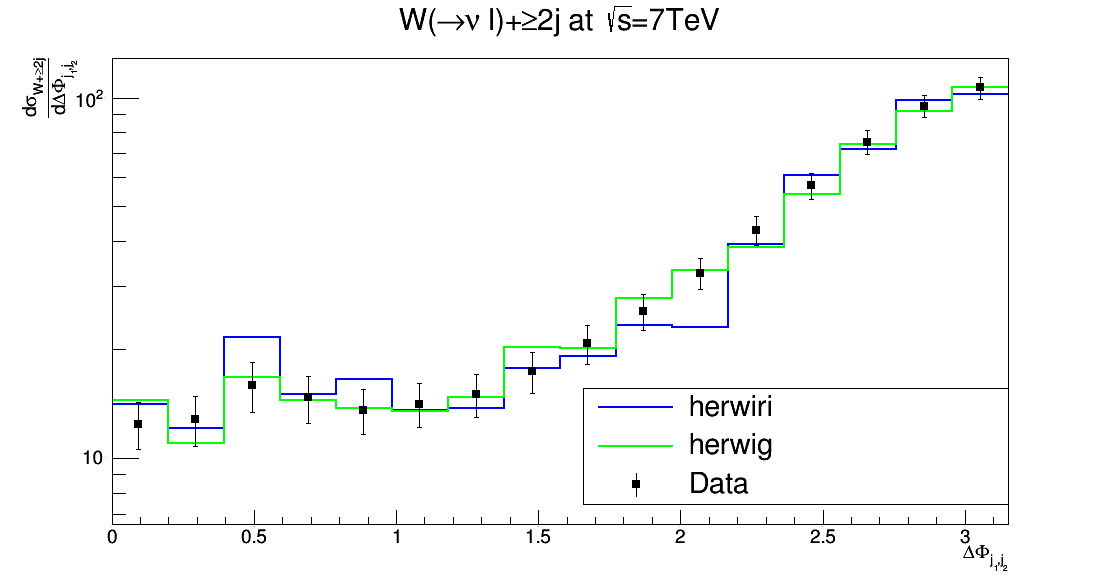}
\caption{Cross section for the production of W~+ jets as a function of the difference in the azimuthal angle between the two leading jets in $N_{jet}\geq 2.$ The data are compared to predictions from MADGRAPH5\_aMC@NLO/HERWIRI1.031 and MADGRAPH5\_aMC@NLO/HERWIG6.521.}
\label{fig13}
\end{figure}

The differential cross sections for the production of W~+~$\geq$2 jets as a function of the azimuthal angle between the two leading jets are shown in Figure.~\ref{fig13}. For $\Delta\phi_{j_{1},j_{2}}< 0.4$, $1<\Delta\phi_{j_{1},j_{2}}<1.4$, and $\Delta\phi_{j_{1},j_{2}}>2.2$, the predicted cross sections by HERWIRI and HERWIG are within the error bars on the data: $\big(\frac{\chi^2}{d.o.f}\big)_{\texttt{HERWIRI}}=1.46$ and $\big(\frac{\chi^2}{d.o.f}\big)_{\texttt{HERWIG}}=0.49$, so that, while both predictions give acceptable fits to the data, the HERWIG fit is the better one.
\subsection{Scalar Sum $H_{T}$}
In this subsection we will study the W~+ jets cross sections as a function of $H_{T}$, the summed scalar $P_{T}$ of all identified objects in the final state. For example, for a prototypical process
\begin{equation}
    pp\rightarrow l+\nu_{l}+j_{1}+j_{2},
\end{equation}
we define $H_{T}$ as follows
\begin{equation}
    H_{T}=P_{T}(l)+P_{T}(\nu_{l})+P_{T}(j_{1})+P_{T}(j_{2}),
\end{equation}
where $l={e,\mu}$.
\par
\begin{figure}[H]
\centering
\includegraphics[scale=0.4]{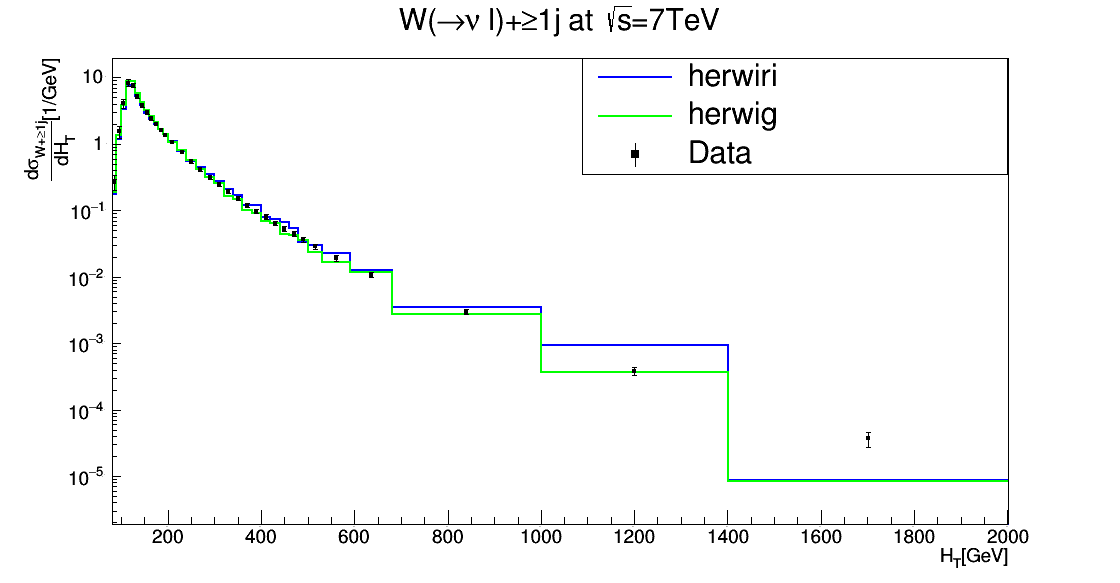}
\caption{Cross section for the production of W~+ jets as a function of the scalar sum $H_{T}$ in $N_{jet}\geq 1.$ The data are compared to predictions from MADGRAPH5\_aMC@NLO/HERWIRI1.031 and MADGRAPH5\_aMC@NLO/HERWIG6.521.}
\label{fig14}
\end{figure}
\begin{figure}[H]
\includegraphics[scale=0.4]{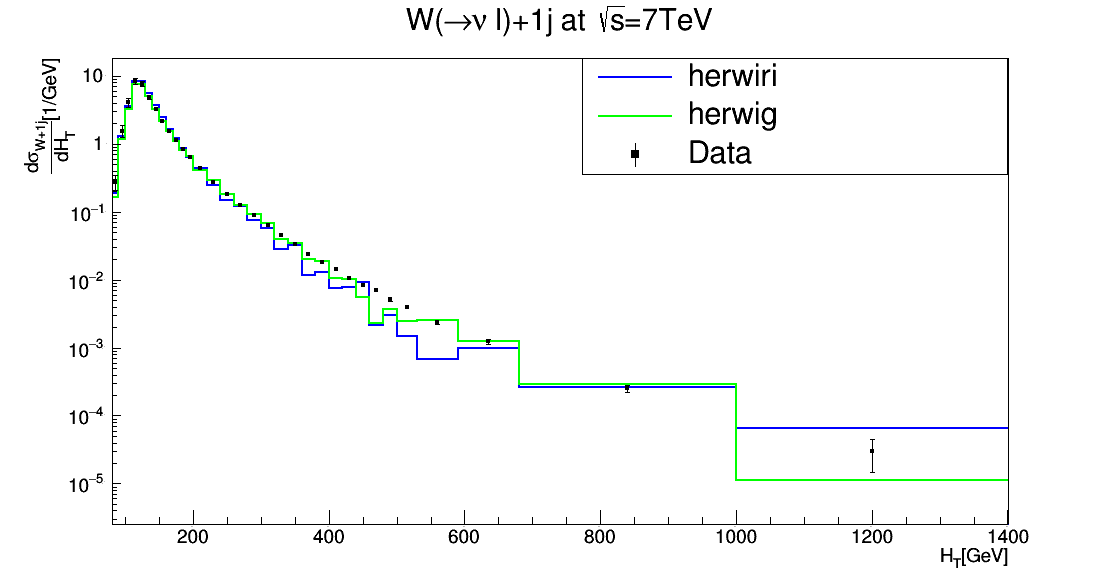}
\caption{Cross section for the production of W~+ jets as a function of the the scalar sum $H_{T}$ in $N_{jet}=1.$ The data are compared to predictions from MADGRAPH5\_aMC@NLO/HERWIRI1.031 and MADGRAPH5\_aMC@NLO/HERWIG6.521.}
\label{fig15}
\end{figure}
\begin{figure}[H]
\centering
\includegraphics[scale=0.4]{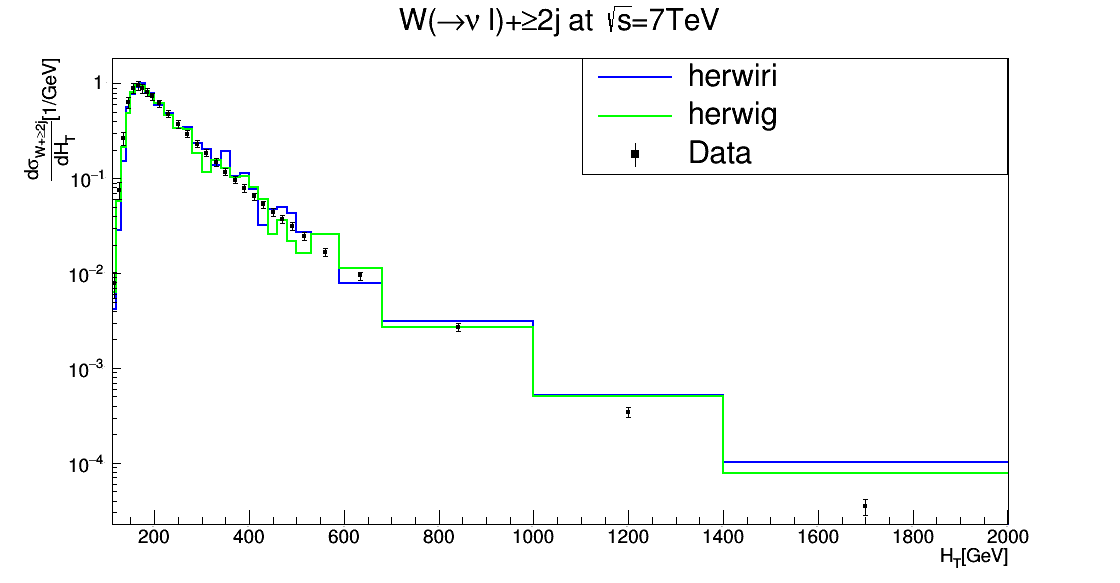}
\caption{Cross section for the production of W~+ jets as a function of the scalar sum $H_{T}$ in $N_{jet}\geq 2.$ The data are compared to predictions from MADGRAPH5\_aMC@NLO/HERWIRI1.031 and MADGRAPH5\_aMC@NLO/HERWIG6.521.}
\label{fig16}
\end{figure}
\begin{figure}[H]
\includegraphics[scale=0.4]{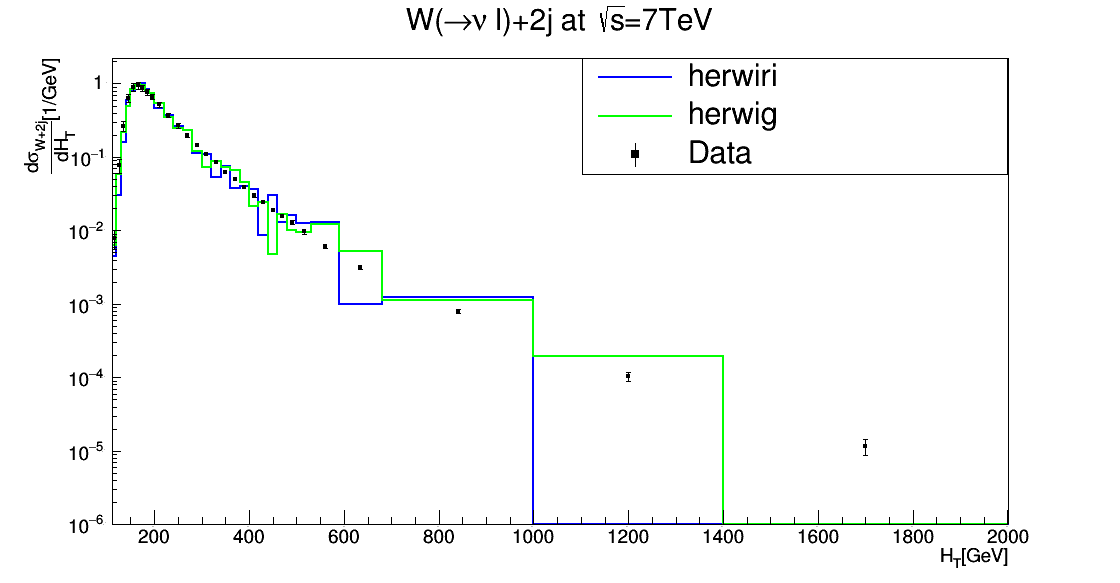}
\caption{Cross section for the production of W~+ jets as a function of the the scalar sum $H_{T}$ in $N_{jet}=2.$ The data are compared to predictions from MADGRAPH5\_aMC@NLO/HERWIRI1.031 and MADGRAPH5\_aMC@NLO/HERWIG6.521.}
\label{fig17}
\end{figure}
\begin{figure}[H]
\centering
\includegraphics[scale=0.4]{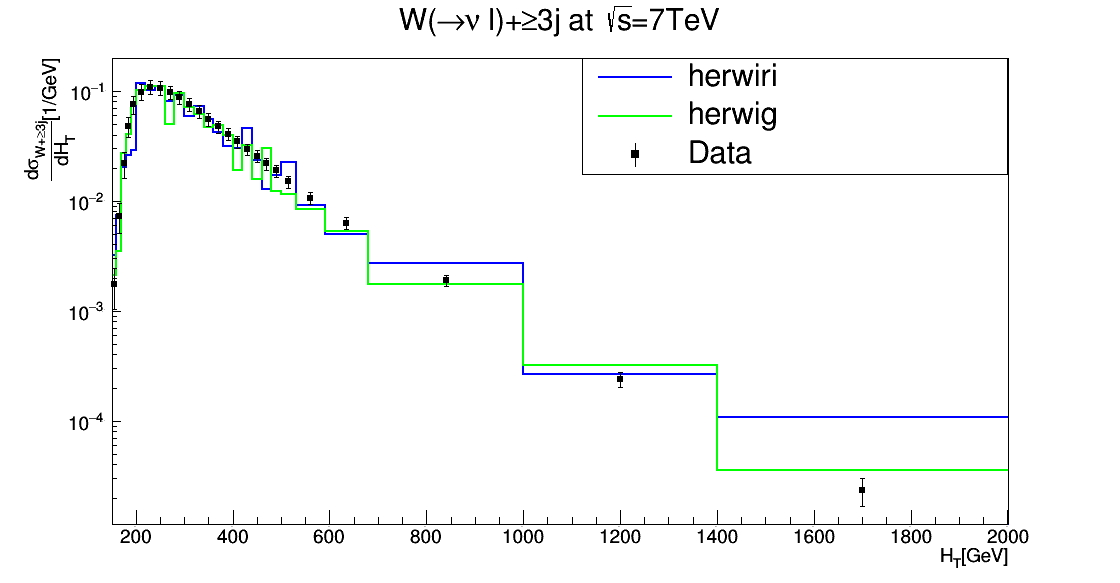}
\caption{Cross section for the production of W~+ jets as a function of the scalar sum $H_{T}$ in $N_{jet}\geq 3.$ The data are compared to predictions from MADGRAPH5\_aMC@NLO/HERWIRI1.031 and MADGRAPH5\_aMC@NLO/HERWIG6.521.}
\label{fig18}
\end{figure}
\begin{figure}[H]
\includegraphics[scale=0.4]{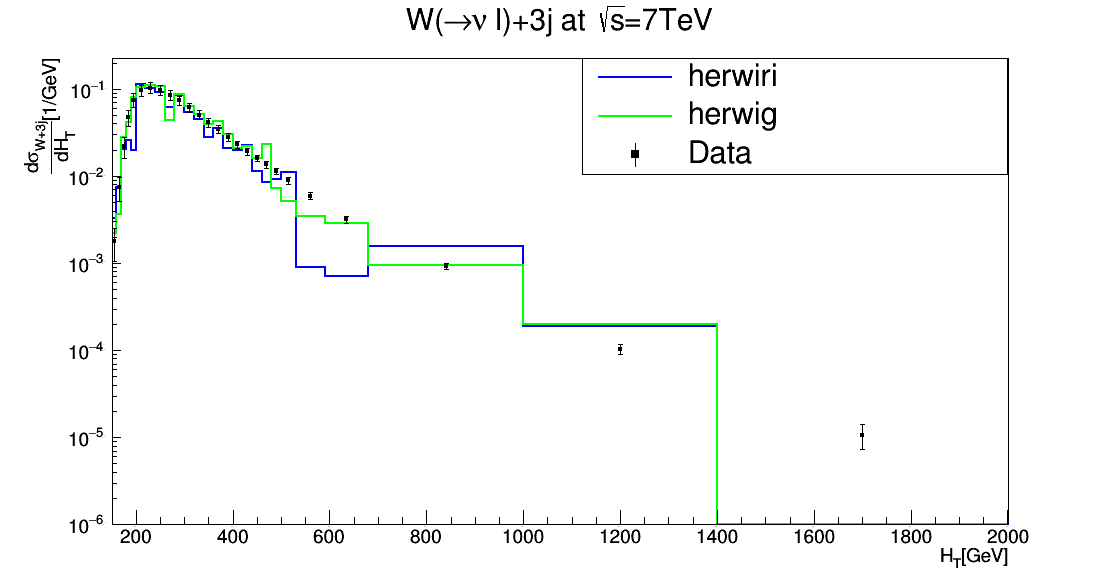}
\caption{Cross section for the production of W~+ jets as a function of the the scalar sum $H_{T}$ in $N_{jet}=3.$ The data are compared to predictions from MADGRAPH5\_aMC@NLO/HERWIRI1.031 and MADGRAPH5\_aMC@NLO/HERWIG6.521.}
\label{fig19}
\end{figure}

The differential cross sections as a function of $H_{T}$ are shown in Figure~\ref{fig14}, Figure~\ref{fig15}, Figure~\ref{fig16}, Figure~\ref{fig17}, Figure~\ref{fig18}, and Figure~\ref{fig19} respectively. We will study the W~+~jets cross sections as a function of $H_{T}$ for low $H_{T}$. We will see in some cases HERWIRI predictions are in agreement with the data and in some cases HERWIG predictions give a better fit to the data. In general, a better agreement is provided for the lower jet multiplicities, e.g. W~+~1 jet and $W+~\geq1$~jet. 
\par
The differential cross sections for the production of W~+~$\geq$1 jet as a function of the scalar sum $H_{T}$ are shown in Figure.~\ref{fig14}. For $H_{T}<300~\mathrm{GeV}$, HERWIRI and HERWIG predictions are in good agreement with data where: $\big(\frac{\chi^2}{d.o.f}\big)_{\texttt{HERWIRI}}=0.591$ and $\big(\frac{\chi^2}{d.o.f}\big)_{\texttt{HERWIG}}=0.96$. For $400<H_{T}<1400~\mathrm{GeV}$, the differential cross sections are fairly well modeled by the HERWIG predictions. (See Appendix C)
\par
The differential cross sections for the production of W~+~1 jet as a function of the scalar sum $H_{T}$ are shown in Figure.~\ref{fig15}. For the case $H_{T}<275~\mathrm{GeV}$, HERWIG predictions are in better  agreement with the data while the predictions provided by HERWIRI either overestimate or underestimate the data in some cases: $\big(\frac{\chi^2}{d.o.f}\big)_{\texttt{HERWIRI}}=3.50$ and $\big(\frac{\chi^2}{d.o.f}\big)_{\texttt{HERWIG}}=0.76$. For $275<H_{T}<1000~\mathrm{GeV}$, the differential cross sections are fairly well modeled by HERWIG predictions. HERWIRI predictions in almost all cases underestimate the data for $275<H_{T}<1000~\mathrm{GeV}$. (See Appendix C)
The differential cross sections for the production of W~+~$\geq$2 jets as a function of the scalar sum $H_{T}$ are shown in Figure.~\ref{fig16}. The predictions provided by HERWIG give a better fit to the data in $H_{T}<275~\mathrm{GeV}$, with $\big(\frac{\chi^2}{d.o.f}\big)_{\texttt{HERWIRI}}=2.25$ and $\big(\frac{\chi^2}{d.o.f}\big)_{\texttt{HERWIG}}=1.26$. In the $275<H_{T}<450~\mathrm{GeV}$ range, HERWIRI gives a better fit to the data; in the $450<H_{T}<650~\mathrm{GeV}$ range, HERWIG predictions are in better agreement with the data. For large $H_{T}$, HERWIG predictions are either in agreement with the data or have less discrepancy with the data than the results provided by HERWIRI, as Figure.~\ref{fig16} reveals.  \par
The differential cross sections for the production of W~+~2 jets as a function of the scalar sum $H_{T}$ are shown in Figure.~\ref{fig17}. HERWIRI and HERWIG seem to be unable to provide a good fit for the data at $H_{T}<190~\mathrm{GeV}$ where they underestimate the data; In the $H_{T}<250~\mathrm{GeV}$ range, HERWIG predictions are in better agreement with the data, where $\big(\frac{\chi^2}{d.o.f}\big)_{\texttt{HERWIRI}}=2.36$ and $\big(\frac{\chi^2}{d.o.f}\big)_{\texttt{HERWIG}}=1.09$. \par
At scalar sum values around $170<H_{T}<250$~GeV, HERWIRI and HERWIG predictions overlap fairly well with the data. In general, we conclude that the discrepancy of the predictions provided by HERWIRI is less than that of HERWIG.

The differential cross sections for the production of W~+~$\geq$3 jets as a function of the scalar sum $H_{T}$ are shown in Figure.~\ref{fig18}. A good fit is provided by the HERWIG predictions for $H_{T}<275~\mathrm{GeV}$, where $\big(\frac{\chi^2}{d.o.f}\big)_{\texttt{HERWIRI}}=2.71$ and $\big(\frac{\chi^2}{d.o.f}\big)_{\texttt{HERWIG}}=2.01$. The HERWIG and HERWIRI predictions overlap fairly well with the data for $275<H_{T}<400~\mathrm{GeV}.$ For the higher range $650<H_{T}<2000~\mathrm{GeV},$ the HERWIG predictions are in better agreement with the data while in most cases HERWIRI either underestimates or overestimates the data.\par 

The differential cross sections for the production of W~+~3 jets as a function of the scalar sum $H_{T}$ are shown in Figure.~\ref{fig19}. HERWIG gives a better fit to the data for $H_{T}<250$, with $\big(\frac{\chi^2}{d.o.f}\big)_{\texttt{HERWIRI}}=3.73$ and $\big(\frac{\chi^2}{d.o.f}\big)_{\texttt{HERWIG}}=0.79$. In general, the predictions provided by HERWIG give a better fit to the data. 
\subsection{Scalar Sum $S_{T}$}
\begin{figure}[H]
\centering
\includegraphics[scale=0.4]{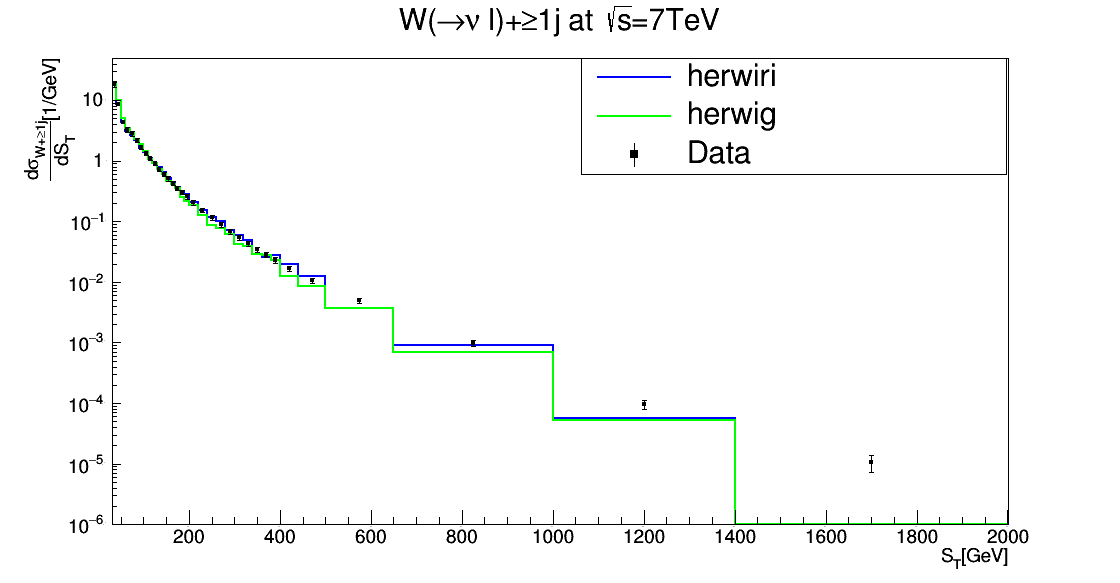}
\caption{Cross section for the production of W~+ jets as a function of the scalar sum $S_{T}$ in $N_{jet}\geq 1.$ The data are compared to predictions from MADGRAPH5\_aMC@NLO/HERWIRI1.031 and MADGRAPH5\_aMC@NLO/HERWIG6.521.}
\label{fig20}
\end{figure}

\begin{figure}[H]
\centering
\includegraphics[scale=0.4]{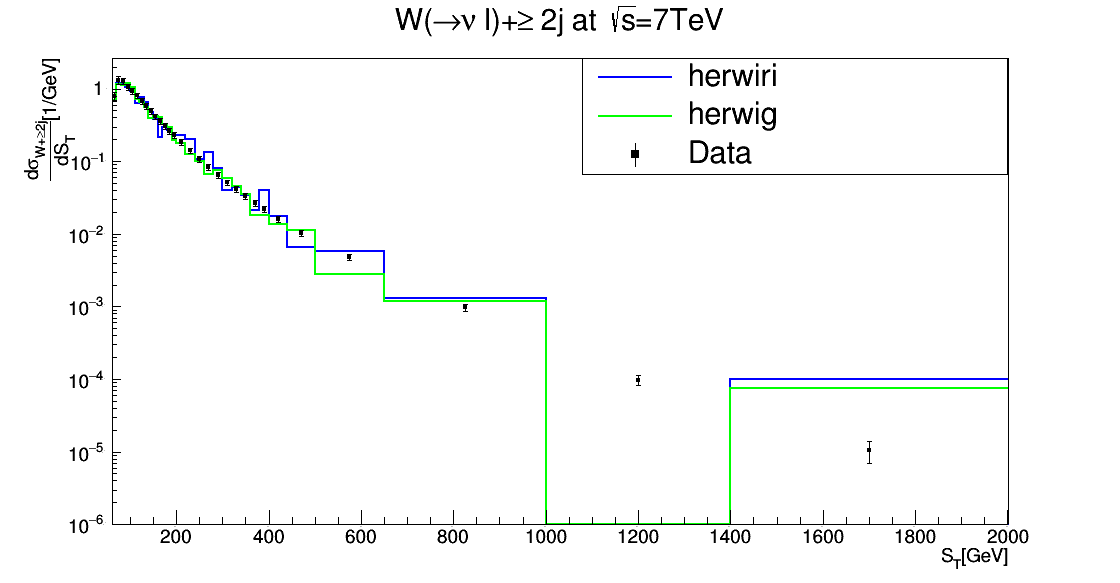}
\caption{Cross section for the production of W~+ jets as a function of the scalar sum $S_{T}$ in $N_{jet}\geq 2.$ The data are compared to predictions from MADGRAPH5\_aMC@NLO/HERWIRI1.031 and MADGRAPH5\_aMC@NLO/HERWIG6.521.}
\label{fig21}
\end{figure}

\begin{figure}[H]
\includegraphics[scale=0.4]{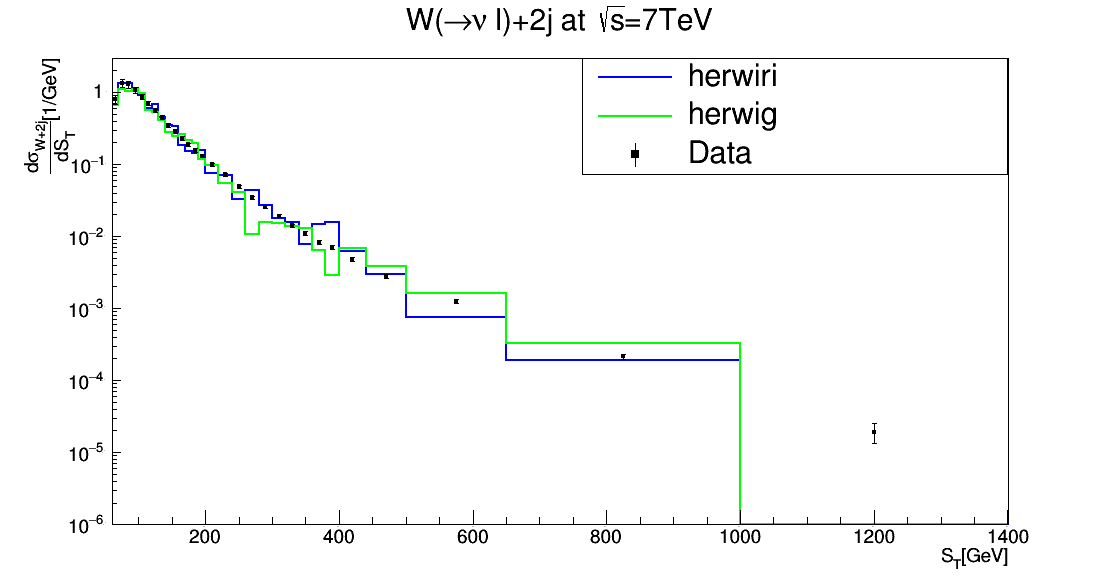}
\caption{Cross section for the production of W~+ jets as a function of the the scalar sum $S_{T}$ in $N_{jet}=2.$ The data are compared to predictions from MADGRAPH5\_aMC@NLO/HERWIRI1.031 and MADGRAPH5\_aMC@NLO/HERWIG6.521.}
\label{fig22}
\end{figure}

\begin{figure}[H]
\centering
\includegraphics[scale=0.4]{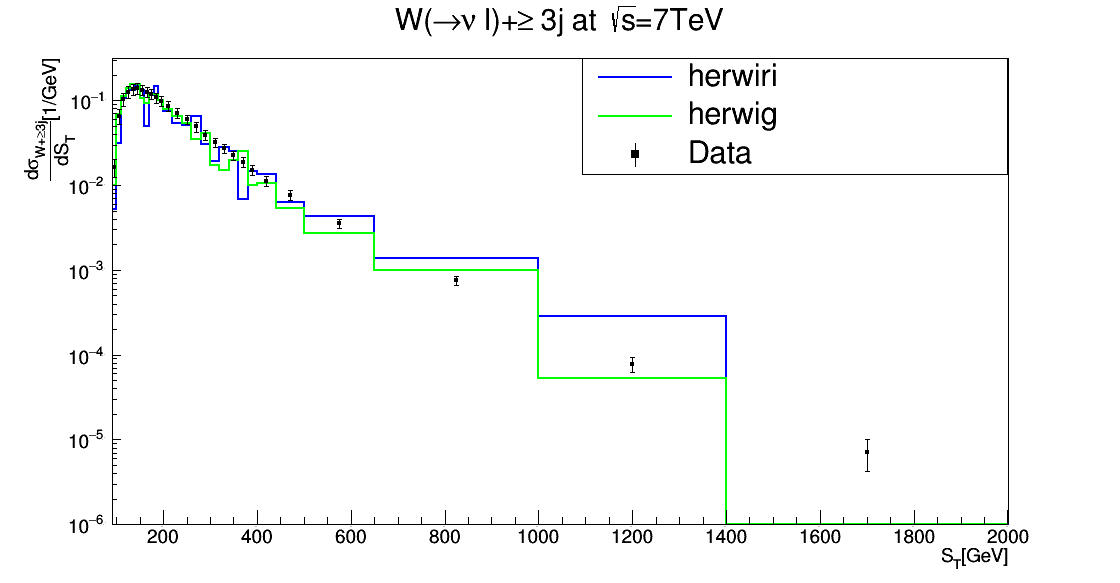}
\caption{Cross section for the production of W~+ jets as a function of the scalar sum $S_{T}$ in $N_{jet}\geq 3.$ The data are compared to predictions from MADGRAPH5\_aMC@NLO/HERWIRI1.031 and MADGRAPH5\_aMC@NLO/HERWIG6.521.}
\label{fig23}
\end{figure}

\begin{figure}[H]
\includegraphics[scale=0.4]{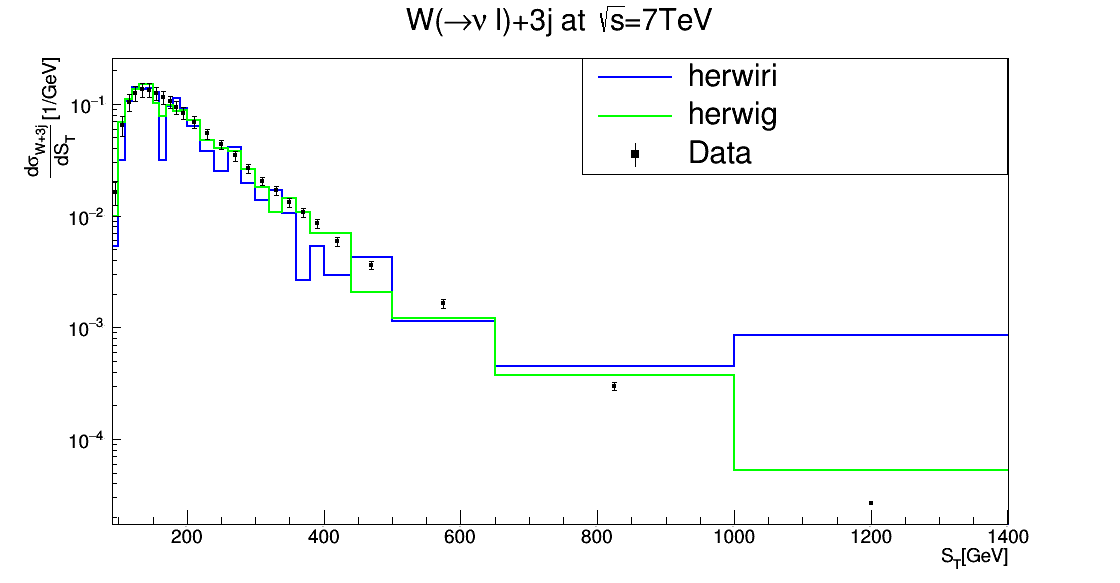}
\caption{Cross section for the production of W~+ jets as a function of the the scalar sum $S_{T}$ in $N_{jet}=3.$ The data are compared to predictions from MADGRAPH5\_aMC@NLO/HERWIRI1.031 and MADGRAPH5\_aMC@NLO/HERWIG6.521.}
\label{fig24}
\end{figure}

In this subsection, we study the behavior of W~+ jets cross sections as a function of the scalar sum $S_{T}$, where $S_{T}$ is defined as the summed scalar $P_{T}$ of all the jets in the event:
\begin{equation}
    S_{T}=\sum_{i=1 }^\mathit{Njet}|P_{T}(i)|,
\end{equation}
where $|P_{T}(i)|$ is the transverse momentum of the $i$th jet and \textit{Njet} is the maximum number of jets in each event. 
The differential cross sections as a function of $S_{T}$ are shown in Figure.~\ref{fig20}, Figure.~\ref{fig21}, Figure.~\ref{fig22}, Figure.~\ref{fig23}, and Figure.~\ref{fig24} respectively. We will study the W~+~jets cross sections as a function of $S_{T}$ for low $S_{T}$. We will see in some cases HERWIRI predictions are in agreement with the data and in some cases HERWIG predictions give a better fit to the data. In general, a better agreement is provided for the lower jet multiplicities, e.g. W~+~1 jet and $W+~\geq1$~jet. 

The differential cross sections for the production of W~+~$\geq$1 jet as a function of the scalar sum $S_{T}$ are shown in Figure.~\ref{fig20}. A good fit to the data is provided by HERWIRI at $S_{T}<300~\mathrm{GeV}$ while HERWIG predictions lie below the data in some cases: $\big(\frac{\chi^2}{d.o.f}\big)_{\texttt{HERWIRI}}=0.28$ and $\big(\frac{\chi^2}{d.o.f}\big)_{\texttt{HERWIG}}=1.94$. For $300<S_{T}<1000~\mathrm{GeV},$ the HERWIRI predictions are in good agreement with the data. For higher values of $S_{T}$, $1000<S_{T}<2000~\mathrm{GeV}$, HERWIRI and HERWIG predictions underestimate the data. \par
The differential cross sections for the production of W~+~$\geq$2 jets as a function of the scalar sum $S_{T}$ are shown in Figure.~\ref{fig21}. For $S_{T}<200~\mathrm{GeV},$~the predictions provided by HERWIG are in better agreement with the data: $\big(\frac{\chi^2}{d.o.f}\big)_{\texttt{HERWIRI}}=2.96$ and $\big(\frac{\chi^2}{d.o.f}\big)_{\texttt{HERWIG}}=1.65$. For medium values of $S_{T}$, the HERWIG predictions give a fair fit to the data.  For large $S_{T}$ values, in some cases HERWIG gives a better fit to the data. \par
The differential cross sections for the production of W~+~2 jets as a function of the scalar sum $S_{T}$ are shown in Figure.~\ref{fig22}. Good agreement is provided by the predictions of HERWIG for $S_{T}<200~\mathrm{GeV}$, where $\big(\frac{\chi^2}{d.o.f}\big)_{\texttt{HERWIRI}}=4.39$ and $\big(\frac{\chi^2}{d.o.f}\big)_{\texttt{HERWIG}}=5.27$. HERWIRI in general gives either a better fit to the data or less discrepancy in comparison with HERWIG. \par
The differential cross sections for the production of W~+~$\geq$3 jets as a function of the scalar sum $S_{T}$ are shown in Figure.~\ref{fig23}. For $S_{T}<200~\mathrm{GeV}$, the predictions provided by HERWIG give a better fit to the data where $\big(\frac{\chi^2}{d.o.f}\big)_{\texttt{HERWIRI}}=3.80$ and $\big(\frac{\chi^2}{d.o.f}\big)_{\texttt{HERWIG}}=1.05$.
\par
The differential cross sections for the production of W~+~3 jets as a function of the scalar sum $S_{T}$ are shown in Figure.~\ref{fig24}. For $S_{T}<200~\mathrm{GeV}$, the predictions provided by HERWIG give a better fit to the data, with $\big(\frac{\chi^2}{d.o.f}\big)_{\texttt{HERWIRI}}=4.54$ and $\big(\frac{\chi^2}{d.o.f}\big)_{\texttt{HERWIG}}=1.30$.
\par
It is clear in some cases HERWIRI predictions are in agreement with the data and in some cases HERWIG predictions give a better fit to the data. In general, a better agreement is provided for the lower jet multiplicities, e.g. W~+~1 jet and $W+~\geq1$~jet. 

\subsection{Cross Sections}
\begin{figure}[H]
\centering
\includegraphics[scale=0.4]{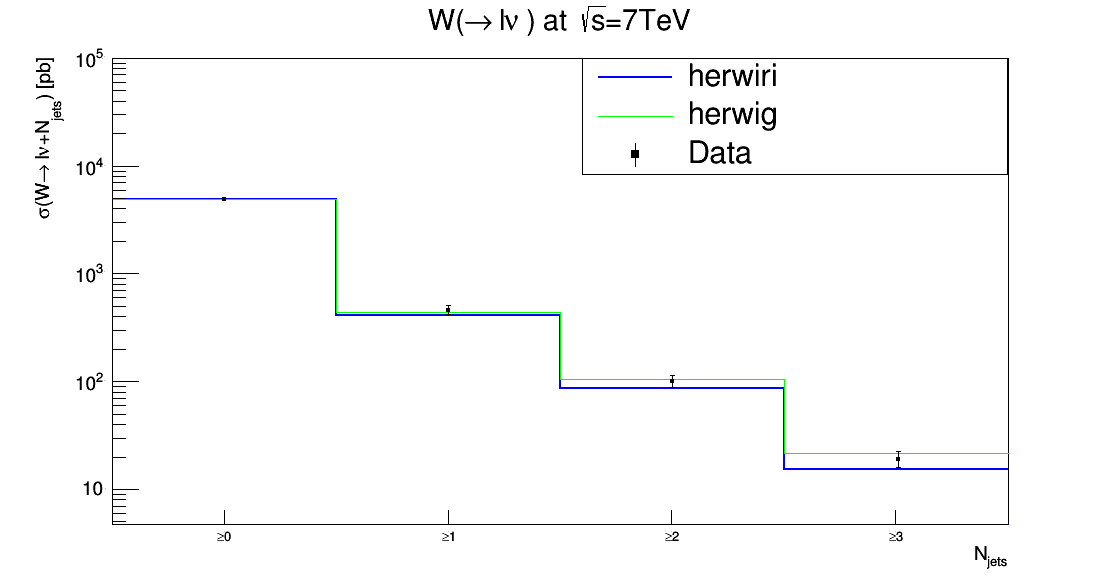}
\caption{Cross section for the production of W~+ jets as a function of the inclusive jet multiplicity. The data are compared to predictions from MADGRAPH5\_aMC@NLO/HERWIRI1.031 and MADGRAPH5\_aMC@NLO/HERWIG6.521}
\label{fig25}
\end{figure}

\begin{figure}[H]
\includegraphics[scale=0.4]{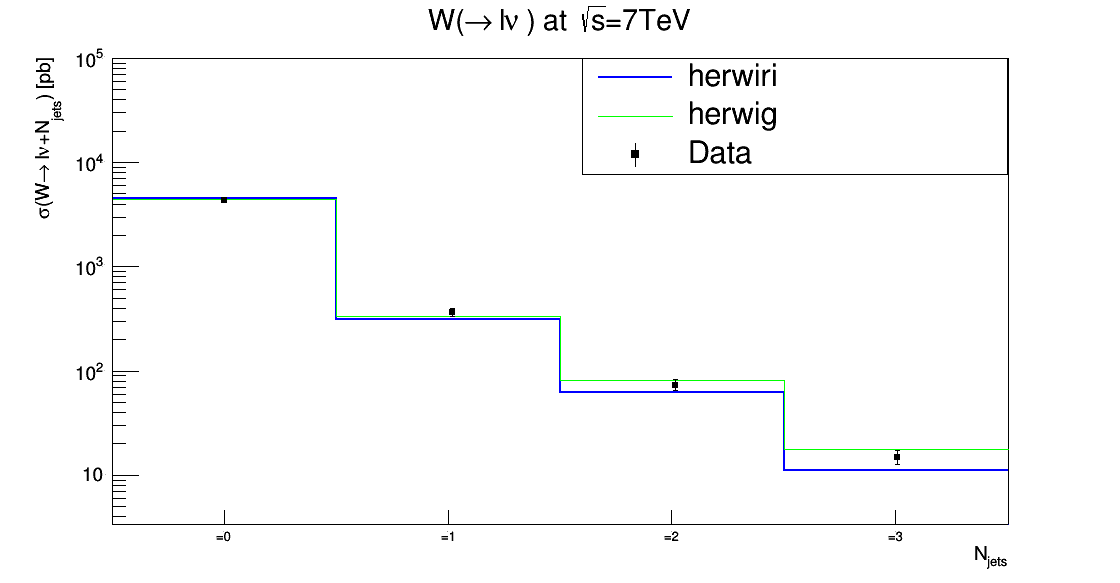}
\caption{Cross section for the production of W~+ jets as a function of the exclusive jet multiplicity. The data are compared to predictions from MADGRAPH5\_aMC@NLO/HERWIRI1.031 and MADGRAPH5\_aMC@NLO/HERWIG6.521}
\label{fig26}
\end{figure}

The cross sections for $W\rightarrow l+\nu_{l}$ production as functions of the inclusive and exclusive jet multiplicity are shown in Figure.~\ref{fig25} and Figure.~\ref{fig26}.
Figure.~\ref{fig25} shows the cross sections for the production of W~+ jet as a function of the inclusive jet multiplicity. A good fit is provided by HERWIRI and HERWIG for $N_{jet}\geq1$, for $N_{jet}\geq2$ and for $N_{jet}\geq$~3, where the HERWIRI prediction is just at edge of the lower error bar on the data. For the exclusive case in Fig.~\ref{fig26},
similar comments apply except that for the $N_{jet} =$~3 case the HERWIRI prediction is about 2 $\sigma$ below the data. 
\section{Results (CMS Collaboration)}
In this Section the measured W$(\rightarrow\mu+\nu_{\mu})$~+ jets fiducial cross sections \cite{Khachatryan:2014uva} are shown and compared to the predictions of MADGRAPH5\_aMC@NLO/HERWIRI1.031 and MADGRAPH5\_aMC@NLO/HERWIG6.521,which are hereafter oftentimes referred to as HERWIRI and HERWIG, respectively. Each distribution is combined separately by minimizing a $\chi^2$ function. The factors applied to the theory predictions are summarized in Appendix~B.
\subsection{Transverse Momentum Distributions $P_{T}$ }
\begin{figure}[H]
\centering
\includegraphics[scale=0.4]{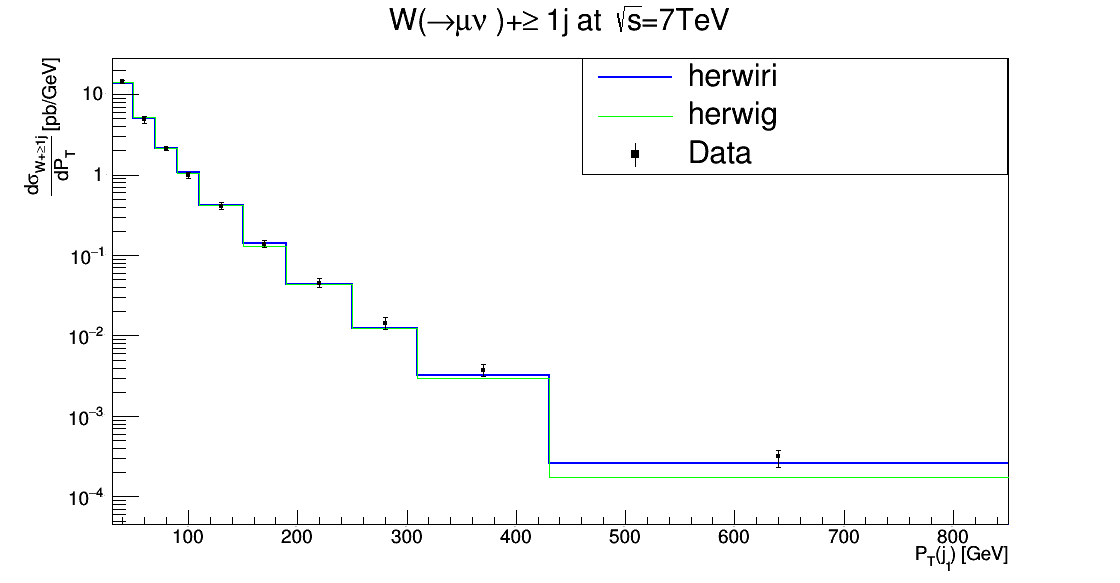}
\caption[Cross section for the production of W~+ jets as a function of the leading jet $P_{T}$ for $N_{jet}\geq 1.$]{Cross section for the production of W~+ jets as a function of the leading jet $P_{T}$ for $N_{jet}\geq 1.$ The data are compared to predictions from MADGRAPH5\_aMC@NLO/HERWIRI1.031 and MADGRAPH5\_aMC@NLO/HERWIG6.521.}
\label{fig1c}
\end{figure}

\begin{figure}[H]
\centering
\includegraphics[scale=0.4]{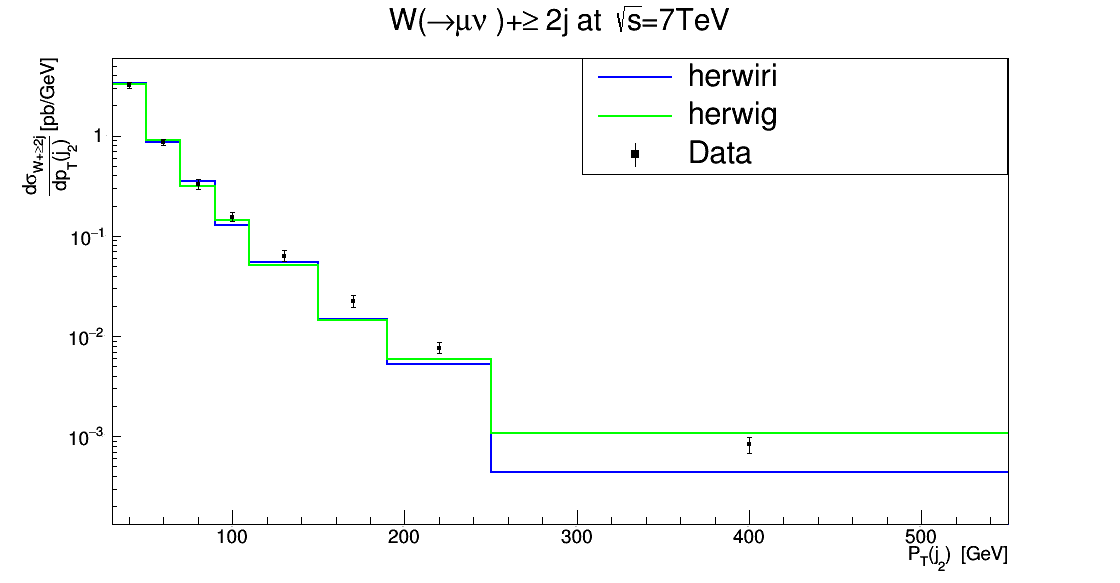}
\caption[Cross section for the production of W~+ jets as a function of the second leading jet $P_{T}$ for $N_{jet}\geq2.$]{Cross section for the production of W~+ jets as a function of the second leading jet $P_{T}$ for $N_{jet}\geq2.$ The data are compared to predictions from MADGRAPH5\_aMC@NLO/HERWIRI1.031 and MADGRAPH5\_aMC@NLO/HERWIG6.521.}
\label{fig2c}
\end{figure}

\begin{figure}[H]
\centering
\includegraphics[scale=0.4]{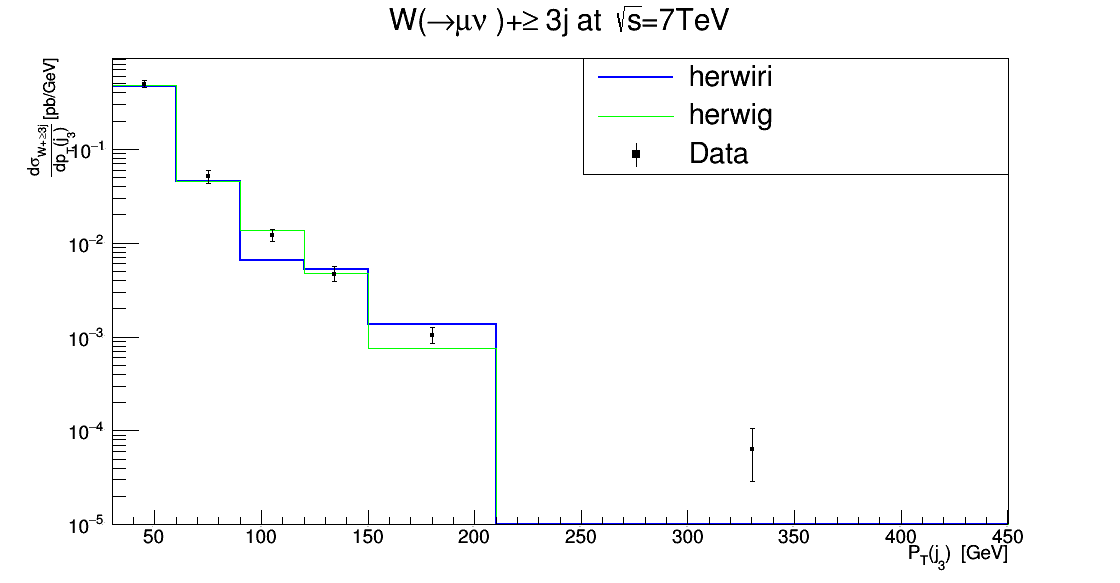}
\caption[Cross section for the production of W~+ jets as a function of the third leading jet $P_{T}$ for $N_{jet}\geq 3.$]{Cross section for the production of W~+ jets as a function of the third leading jet $P_{T}$ for $N_{jet}\geq 3.$ The data are compared to predictions from MADGRAPH5\_aMC@NLO/HERWIRI1.031 and MADGRAPH5\_aMC@NLO/HERWIG6.521.}
\label{fig3c}
\end{figure}

The differential cross sections in jet $P_{T}$ for inclusive jet multiplicities from 1 to 3 are shown in Figure.~\ref{fig1c}, Figure.~\ref{fig2c} and Figure.~\ref{fig3c}, and compared with predictions provided by HERWIRI and HERWIG.
The differential cross sections as functions of the first three leading jets are shown in Fig.~\ref{fig1c}, Fig.~\ref{fig2c}, and Fig.~\ref{fig3c}. In Figure~\ref{fig1c}, for $P_{T}<150~\mathrm{GeV}$, the predictions provided by HERWIRI and HERWIG give a very good fit to the data, with $\big(\frac{\chi^2}{d.o.f}\big)_{\texttt{HERWIRI}}=0.64$ and $\big(\frac{\chi^2}{d.o.f}\big)_{\texttt{HERWIG}}=0.35$. 
\par
In Figure.~\ref{fig2c}, for $P_{T}<110~\mathrm{GeV}$, a better fit is provided by HERWIG  to the data points,  where $\big(\frac{\chi^2}{d.o.f}\big)_{\texttt{HERWIRI}}=1.43$ and $\big(\frac{\chi^2}{d.o.f}\big)_{\texttt{HERWIG}}=0.73$. For higher values of $P_{T}$, the predictions provided by HERWIRI lie below the data while the HERWIG results either underestimate or overestimate the data.
\par
In Figure.~\ref{fig3c}, for $P_{T}<150~\mathrm{GeV}$, the HERWIG predictions, in general, give a better fit to the data:  $\big(\frac{\chi^2}{d.o.f}\big)_{\texttt{HERWIRI}}=2.60$ and $\big(\frac{\chi^2}{d.o.f}\big)_{\texttt{HERWIG}}=1.59$.
\subsection{The Scalar Sum of Jet Transverse Momenta $H_{T}$ }
In this subsection, the differential cross sections are shown as function of $H_{T}$ for inclusive jet multiplicities 1--3. The scalar sum $H_{T}$ is defined as
\begin{equation}
H_{T}=\sum_{i=1}^{N_{jet}}P_{T}(j_{i}),
\end{equation}
for each event.
\par
\begin{figure}[H]
\centering
\includegraphics[scale=0.4]{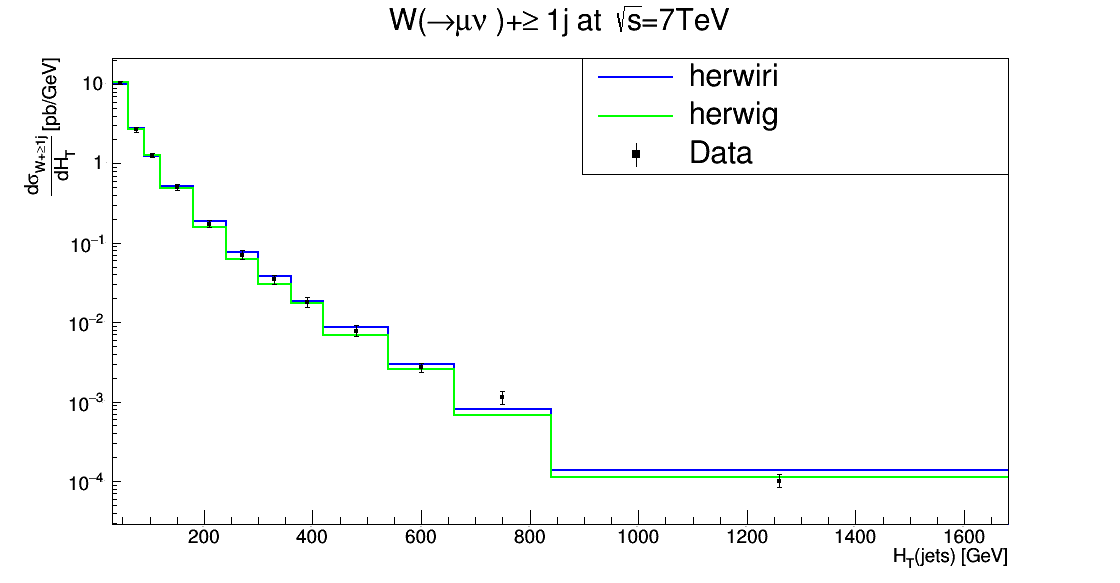}
\caption[Cross section for the production of W~+ jets as a function of  $H_{T}$ for $N_{jet}\geq 1.$]{Cross section for the production of W~+ jets as a function of  $H_{T}$ for $N_{jet}\geq 1.$ The data are compared to predictions from MADGRAPH5\_aMC@NLO/HERWIRI1.031 and MADGRAPH5\_aMC@NLO/HERWIG6.521.}
\label{fig4c}
\end{figure}
\begin{figure}[H]
\centering
\includegraphics[scale=0.4]{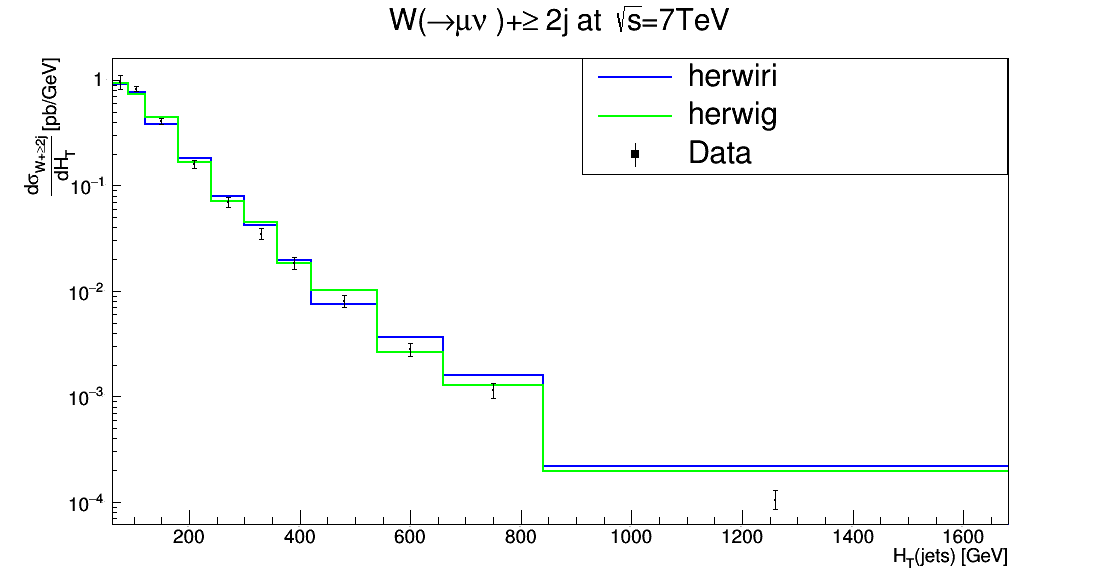}
\caption[Cross section for the production of W~+ jets as a function of  $H_{T}$ for $N_{jet}\geq2.$]{Cross section for the production of W~+ jets as a function of  $H_{T}$ for $N_{jet}\geq2.$ The data are compared to predictions from MADGRAPH5\_aMC@NLO/HERWIRI1.031 and MADGRAPH5\_aMC@NLO/HERWIG6.521.}
\label{fig5c}
\end{figure}
\begin{figure}[H]
\centering
\includegraphics[scale=0.4]{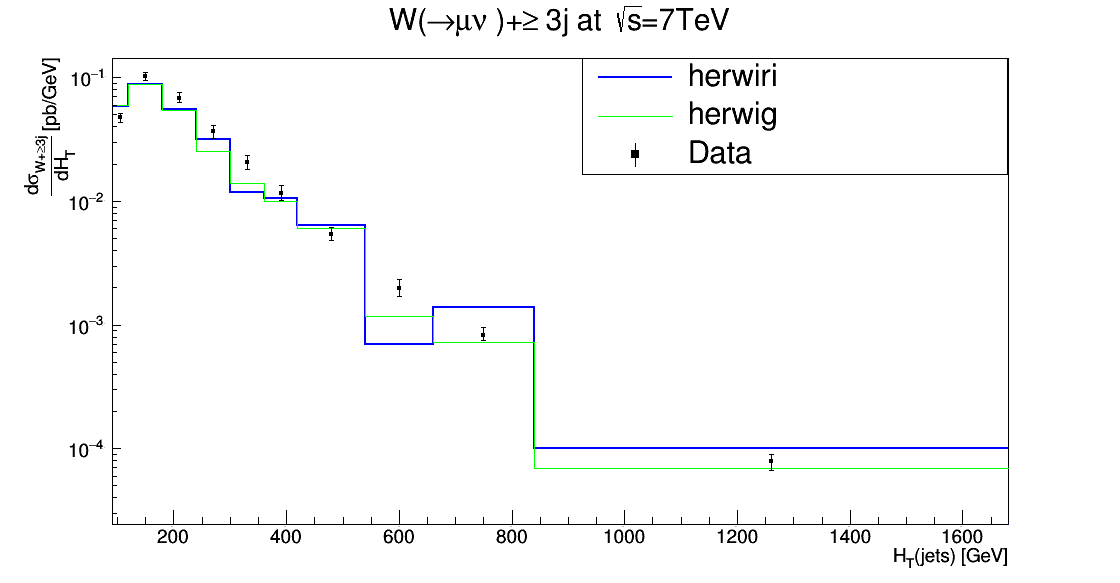}
\caption[Cross section for the production of W~+ jets as a function of $H_{T}$ for $N_{jet}\geq 3.$ ]{Cross section for the production of W~+ jets as a function of $H_{T}$ for $N_{jet}\geq 3.$ The data are compared to predictions from MADGRAPH5\_aMC@NLO/HERWIRI1.031 and MADGRAPH5\_aMC@NLO/HERWIG6.521.}
\label{fig6c}
\end{figure}

The differential cross sections as a function of $H_{T}$ for inclusive jet multiplicities 1--3 are shown in Figure.~\ref{fig4c}, Figure.~\ref{fig5c}, and Figure.~\ref{fig6c}. In Figure.~\ref{fig4c},  for $H_{T}<300~\mathrm{GeV}$, the predictions provided by HERWIRI and HERWIG give a very good fit to the data with $\big(\frac{\chi^2}{d.o.f}\big)_{\texttt{HERWIRI}}=0.57$ and $\big(\frac{\chi^2}{d.o.f}\big)_{\texttt{HERWIG}}=0.40$. In Figure.~\ref{fig5c}, for $H_{T}<180~\mathrm{GeV}$, and $360<H_{T}<540~\mathrm{GeV}$, HERWIRI gives a better fit to the data while in Figure.~\ref{fig6c} the predictions provided by HERWIRI give a better fit to the data for $H_{T}<250~\mathrm{GeV}$. In Figure.~\ref{fig5c},  for $H_{T}<300~\mathrm{GeV}$, $\big(\frac{\chi^2}{d.o.f}\big)_{\texttt{HERWIRI}}=1.70$ and $\big(\frac{\chi^2}{d.o.f}\big)_{\texttt{HERWIG}}=1.36$. In Figure.~\ref{fig6c}, for $H_{T}<250~~\mathrm{GeV}$ $\big(\frac{\chi^2}{d.o.f}\big)_{\texttt{HERWIRI}}=4.02$ and $\big(\frac{\chi^2}{d.o.f}\big)_{\texttt{HERWIG}}=4.37$.


\subsection{Pseudorapidity Distributions $|\eta(j)|$}
In this section, the differential cross sections are shown as functions of pseudorapidities of the three leading jets. The pseudorapidity, which was defined in Eq.~(\ref{etadef}), can be written as 
\begin{equation}
\eta=\frac{1}{2}\ln\left(\frac{|\vec{P}|+P_{L}}{|\vec{P}|-P_{L}}\right)=\arctanh\left(\frac{P_{L}}{|\vec{P}|}\right),
\label{etadef2}
\end{equation}
where $P_{L}$ where is the component of the momentum along the beam axis.\par
The problem with rapidity is that it can be hard to measure for highly
relativistic particles. We need the total momentum vector of a particle,
especially at high values of the rapidity where the $z$ component of the momentum is large, and the beam pipe can be in the way of measuring it precisely.
\par
However, there is a way of defining a quantity that is almost the same thing as the rapidity which is much easier to measure than $y$ for highly energetic particles. This leads to the concept of the pseudorapidity $\eta$, wherein we see from Eq.(\ref{etadef2}) that the magnitude of the momentum cancels out of the ratio in the arguments of the logarithm and the arctanh in the equation.
\par 
Hadron colliders measure physical momenta in terms of transverse momentum, $P_{T}$, polar angle in the transverse plane, $\phi$, and pseudorapidity. To obtain Cartesian momenta $(P_{x}, P_{y}, P_{z})$, (with the $z$-axis defined as the beam axis), the following conversions are used:
\vspace{5mm}
\begin{equation}
    \left\{ \begin{array}{ll}
         P_{x}=P_{T}\cos\phi,\\
         P_{y}=P_{T}\sin\phi,\\
         P_{z}=P_{T}\sinh\eta.        
         \end{array} \right. 
\end{equation}

\begin{figure}[H]
\centering
\includegraphics[scale=0.4]{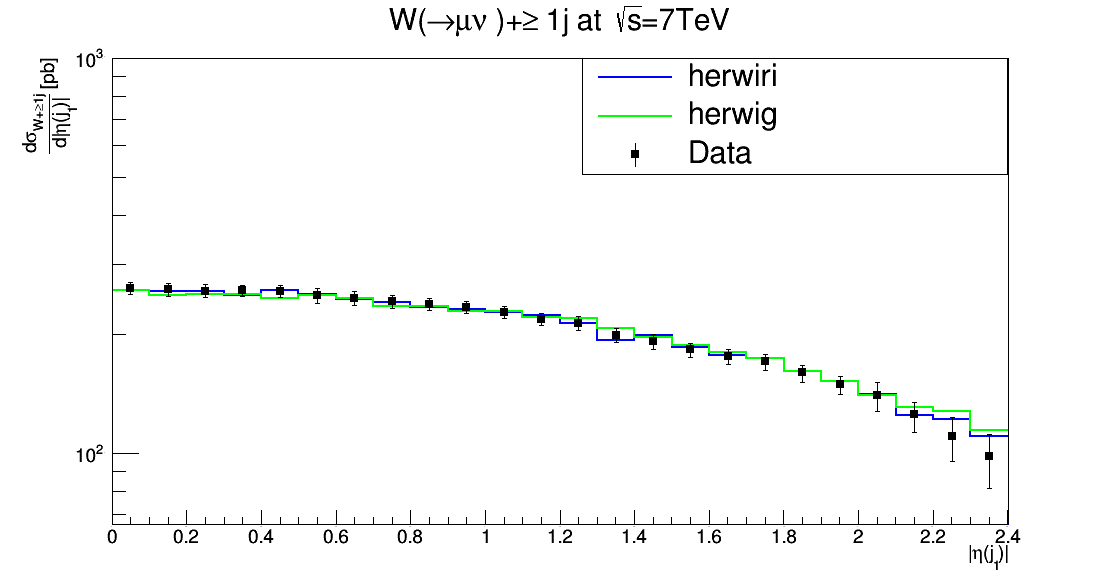}
\caption[Cross section for the production of W~+ jets as a function of  $|\eta(j_{1})|$ for $N_{jet}\geq 1.$]{Cross section for the production of W~+ jets as a function of  $|\eta(j_{1})|$ for $N_{jet}\geq 1.$ The data are compared to predictions from MADGRAPH5\_aMC@NLO/HERWIRI1.031 and MADGRAPH5\_aMC@NLO/HERWIG6.521.}
\label{fig7c}
\end{figure}

\begin{figure}[H]
\centering
\includegraphics[scale=0.4]{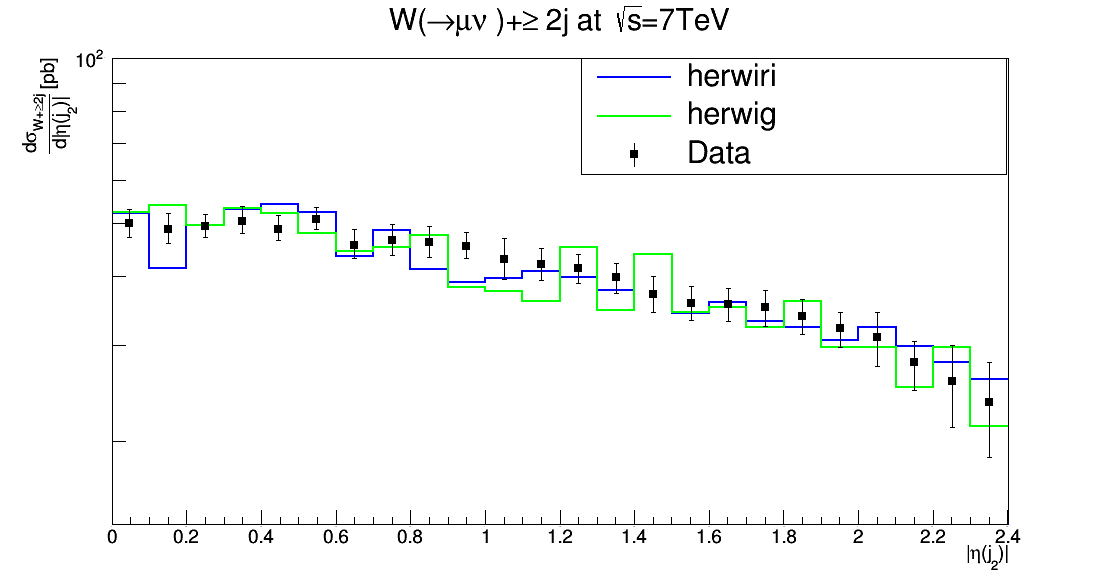}
\caption[Cross section for the production of W~+ jets as a function of  $|\eta(j_{2})|$ for $N_{jet}\geq2.$ ]{Cross section for the production of W~+ jets as a function of  $|\eta(j_{2})|$ for $N_{jet}\geq2.$ The data are compared to predictions from MADGRAPH5\_aMC@NLO/HERWIRI1.031 and MADGRAPH5\_aMC@NLO/HERWIG6.521.}
\label{fig8c}
\end{figure}
\begin{figure}[H]
\centering
\includegraphics[scale=0.4]{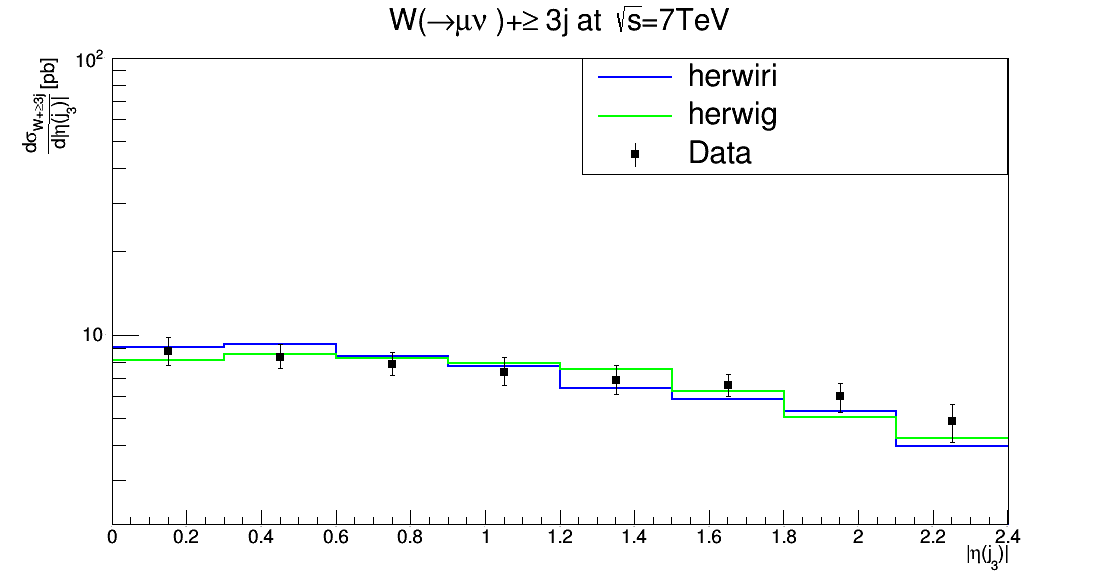}
\caption[Cross section for the production of W~+ jets as a function of $|\eta(j_{1})|$ for $N_{jet}\geq3$.]{Cross section for the production of W~+ jets as a function of $|\eta(j_{1})|$ for $N_{jet}\geq3$. The data are compared to predictions from MADGRAPH5\_aMC@NLO/HERWIRI1.031 and MADGRAPH5\_aMC@NLO/HERWIG6.521.}
\label{fig9c}
\end{figure}
In Figure.~\ref{fig7c} the cross section is shown as a function of $|\eta(j_{1})|$, the leading jet pseudorapidity. The predictions provided by HERWIRI and HERWIG are in good agreement with the data, with $\big(\frac{\chi^2}{d.o.f}\big)_{\texttt{HERWIRI}}=0.39$ and $\big(\frac{\chi^2}{d.o.f}\big)_{\texttt{HERWIG}}=0.79$. In Figure.~\ref{fig8c}, in general, HERWIG gives a better fit to the data, with $\big(\frac{\chi^2}{d.o.f}\big)_{\texttt{HERWIRI}}=1.94$ and $\big(\frac{\chi^2}{d.o.f}\big)_{\texttt{HERWIG}}=1.71$. Figure.~\ref{fig9c} shows that HERWIRI and HERWIG predictions are in agreement with the data, with $\big(\frac{\chi^2}{d.o.f}\big)_{\texttt{HERWIRI}}=0.82$ and $\big(\frac{\chi^2}{d.o.f}\big)_{\texttt{HERWIG}}=0.61$.
\subsection{Azimuthal Angular Distribution Between the Muon and the Leading Jet}
\begin{figure}[t]
\centering
\includegraphics[scale=0.4]{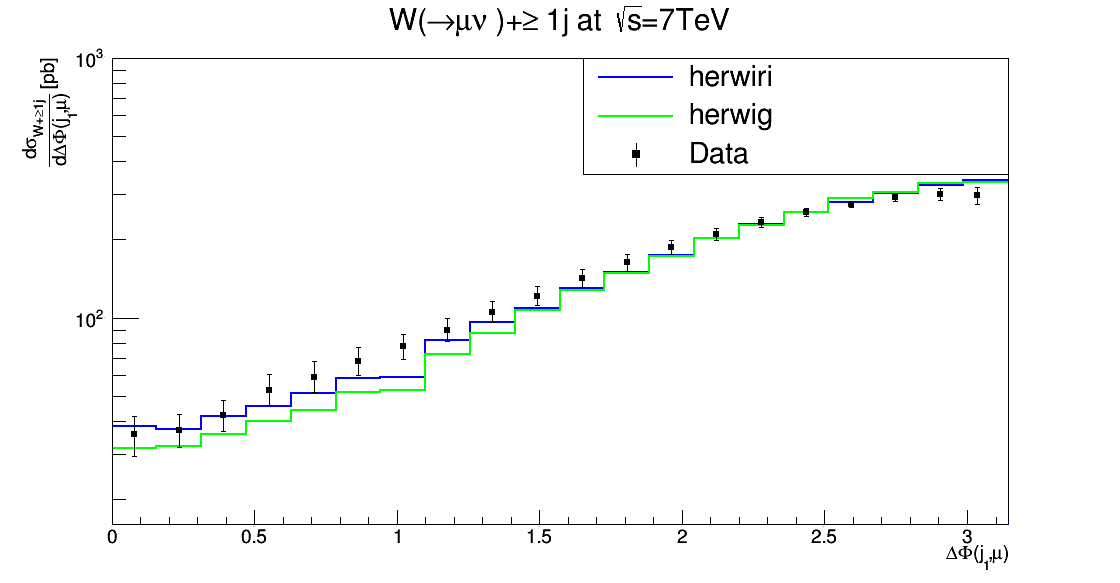}
\caption{Cross section for the production of W~+ jets as a function of the azimuthal angle between the muon and the leading jet $\Delta\Phi(\mu,j_{1})$ for $N_{jet}\geq 1.$ The data are compared to predictions from MADGRAPH5\_aMC@NLO/HERWIRI1.031 and MADGRAPH5\_aMC@NLO/HERWIG6.521.}
\label{fig10c}
\end{figure}

\begin{figure}[H]
\centering
\includegraphics[scale=0.35]{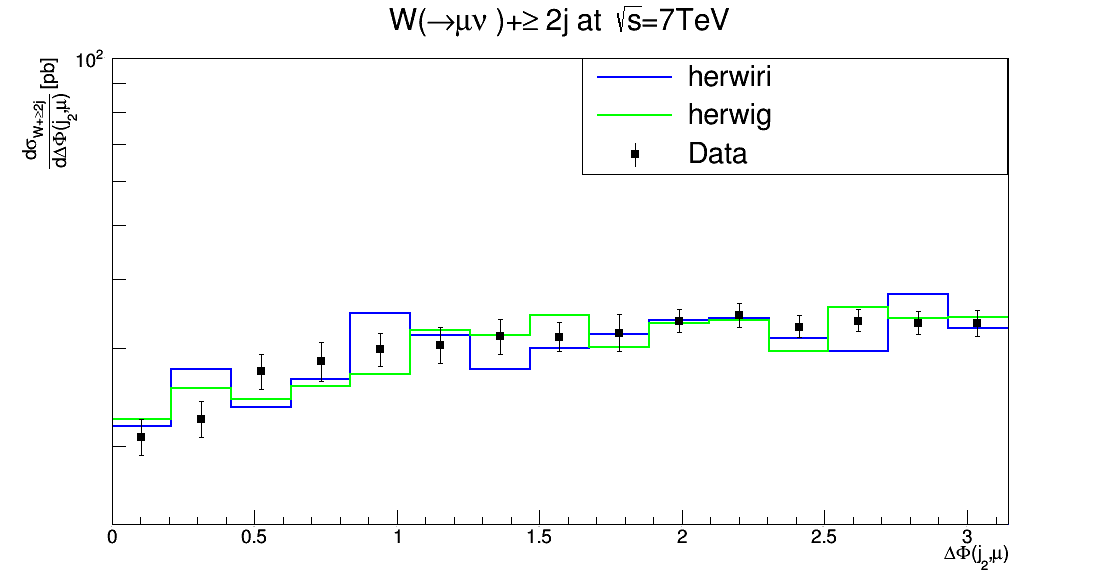}
\caption{Cross section for the production of W~+ jets as a function of  the azimuthal angle between the muon and the second leading jet $\Delta\Phi(\mu,j_{2})$ for $N_{jet}\geq2.$ The data are compared to predictions from MADGRAPH5\_aMC@NLO/HERWIRI1.031 and MADGRAPH5\_aMC@NLO/HERWIG6.521.}
\label{fig11c}
\end{figure}

\begin{figure}[H]
\centering
\includegraphics[scale=0.35]{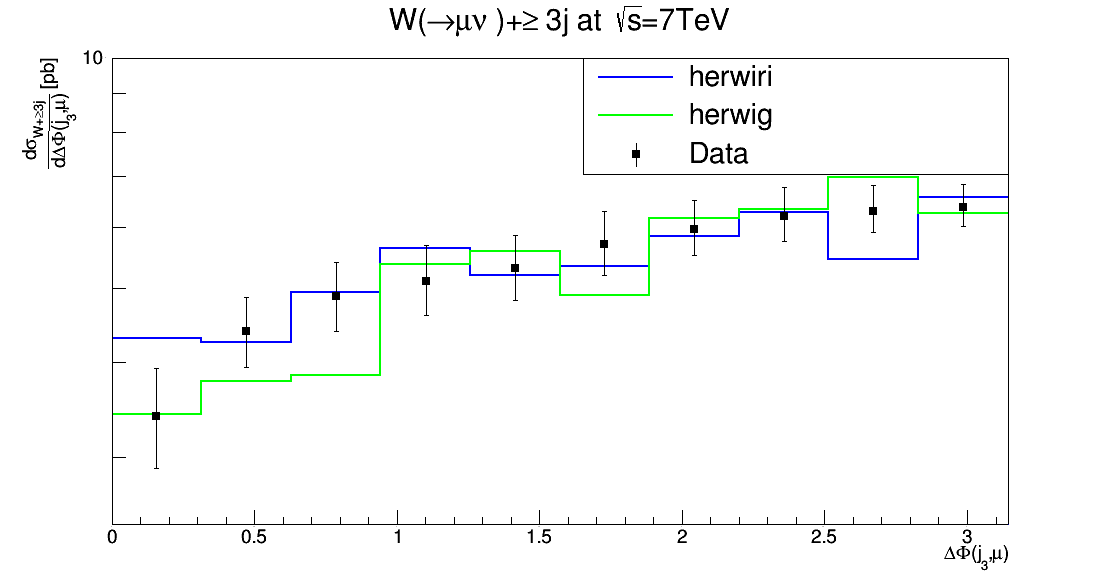}
\caption{Cross section for the production of W~+ jets as a function of the azimuthal angle between the muon and the second leading jet $\Delta\Phi(\mu,j_{3})$ for $N_{jet}\geq 3.$ The data are compared to predictions from MADGRAPH5\_aMC@NLO/HERWIRI1.031 and MADGRAPH5\_aMC@NLO/HERWIG6.521.}
\label{fig12c}
\end{figure}

The differential cross sections are shown as functions of the azimuthal angle between the muon and the first three leading jets for inclusive jet multiplicities 1--3. The azimuthal angle between the muon and the leading jet is defined as
\begin{equation}
\cos(\Delta\Phi(\mu,j_{1}))=\frac{P_{x}(\mu)P_{x}(j_{1})+P_{y}(\mu)P_{y}(j_{1})}{\sqrt{P^2_{x}(\mu)+P^2_{y}(\mu)}\sqrt{P^2_{x}(j_{1})+P^2_{y}(j_{1})}},
\end{equation}
with
\begin{equation}
\left\{ \begin{array}{ll}
         \mu^{\mu}=(E_{\mu},P_{x}(\mu),P_{y}(\mu),P_{L}(\mu)),\\
       j^{\mu}_{1}=(E_{j_{1}},P_{x}(j_{1}),P_{y}(j_{1}),P_{L}(j_{1})),\\\end{array} \right.
\end{equation}
The differential cross sections as functions of the azimuthal angle between the muon and the first three leading jets are shown in Figure.~\ref{fig10c}, Figure.~\ref{fig11c}, and Figure.~\ref{fig12c} for inclusive jet multiplicities 1--3, respectively.
\par
In Figure.~\ref{fig10c}, the data are better modeled by the predictions provided by HERWIRI as expected. Figure.~\ref{fig11c} shows that the HERWIG predictions give a better fit to the data. In Figure.~\ref{fig12c}, the predictions provided by either HERWIRI and HERWIG are in good agreement with the data. In Figure.~\ref{fig10c}, $\big(\frac{\chi^2}{d.o.f}\big)_{\texttt{HERWIRI}}=1.26$ and $\big(\frac{\chi^2}{d.o.f}\big)_{\texttt{HERWIG}}=2.67$. In Figure.~\ref{fig11c}, $\big(\frac{\chi^2}{d.o.f}\big)_{\texttt{HERWIRI}}=2.73$ and $\big(\frac{\chi^2}{d.o.f}\big)_{\texttt{HERWIG}}=1.48$. In Figure.~\ref{fig12c}, $\big(\frac{\chi^2}{d.o.f}\big)_{\texttt{HERWIRI}}=0.89$ and $\big(\frac{\chi^2}{d.o.f}\big)_{\texttt{HERWIG}}=0.61$.

\begin{figure}[H]
\centering
\includegraphics[scale=0.4]{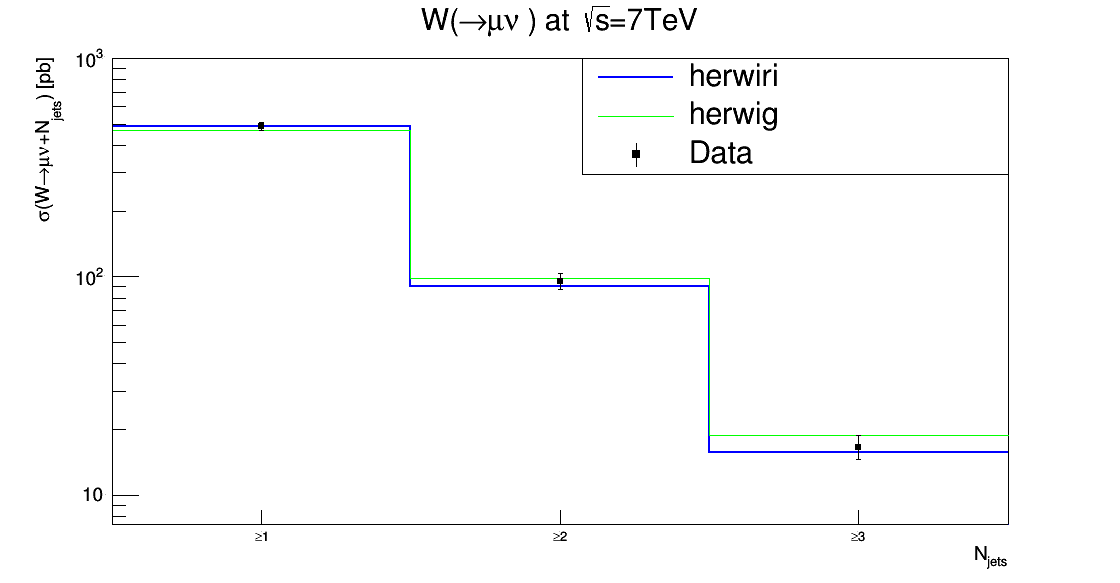}
\caption[Measured cross section versus inclusive jet multiplicity.]{Measured cross section versus inclusive jet multiplicity. The data are compared to predictions from MADGRAPH5\_aMC@NLO/HERWIRI1.031 and MADGRAPH5\_aMC@NLO/HERWIG6.521.}
\label{fig13c}
\end{figure}
\begin{figure}[H]
\centering
\includegraphics[scale=0.4]{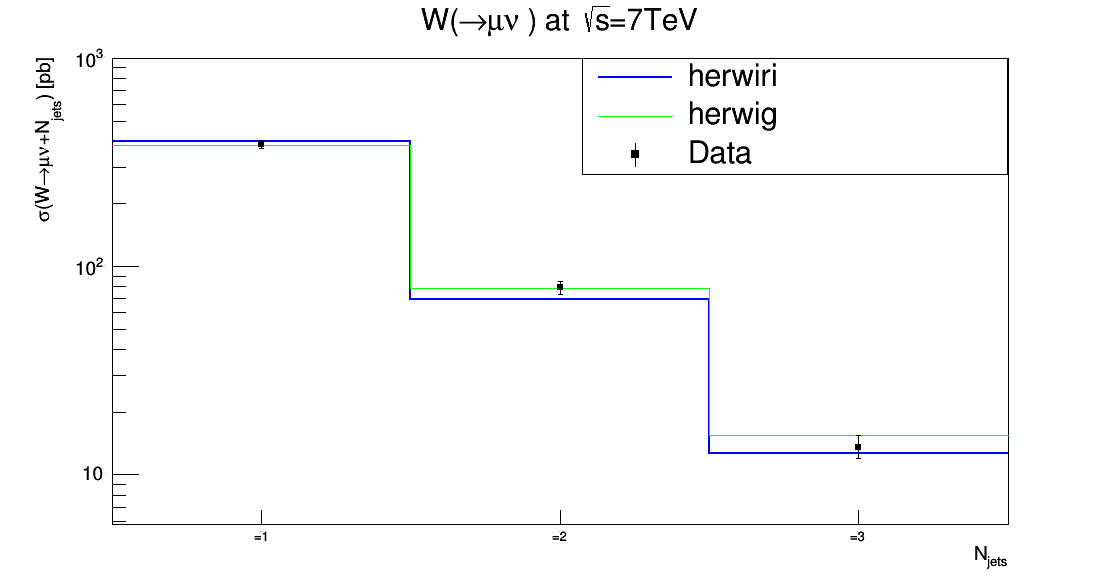}
\caption[Measured cross section versus exclusive jet multiplicity. ]{Measured cross section versus exclusive jet multiplicity. The data are compared to predictions from MADGRAPH5\_aMC@NLO/HERWIRI1.031 and MADGRAPH5\_aMC@NLO/HERWIG6.521.}
\label{fig14c}
\end{figure}
\subsection{Cross Sections}
The measured W$(\rightarrow \mu+\nu_{\mu})$~+~jets fiducial cross sections are shown in Figure.~\ref{fig13c} and Figure.~\ref{fig14c} and compared to the predictions of MADGRAPH5\_aMC@NLO/\\ HERWIRI1.031 and MADGRAPH5\_aMC@NLO/HERWIG6.521.
Figure.~\ref{fig13c} shows the differential cross sections for the inclusive jet multiplicities 1--3. HERWIRI gives a better fit to the data. Figure.~\ref{fig14c} shows the differential cross sections for the exclusive jet multiplicities 1-3. The cross sections provided by HERWIG give a better fit to the data. In Figure.~\ref{fig13c}, $\big(\frac{\chi^2}{d.o.f}\big)_{\texttt{HERWIRI}}=0.46$ and $\big(\frac{\chi^2}{d.o.f}\big)_{\texttt{HERWIG}}=0.56$ while in Figure.~\ref{fig14c}, $\big(\frac{\chi^2}{d.o.f}\big)_{\texttt{HERWIRI}}=1.16$ and $\big(\frac{\chi^2}{d.o.f}\big)_{\texttt{HERWIG}}=0.83$.

\section{Theoretical predictions and associated errors}
Madgraph\_aMC@NLO is only capable of doing the leading-order (LO) and next-to-leading order (NLO) 
calculations. Being that said, the theoretical predictions provided by 
Madgraph\_aMC@NLO would have theoretical errors around 15\%-20\%. For the sake of clarification, four sample plots are given in Appendix D. In the process of generating these sample plots, 20\% theoretical error has been taken into account. (See Figures 49 to 52)

\section{Summary}
The realization of the IR-improved DGLAP-CS theory, when used in the MADGRAPH5\_aMC@NLO/HERWIRI1.031 $\mathcal{O}(\alpha)$ ME-matched parton shower framework, provides us with the opportunity to explain, in the soft regime, the differential cross sections for a W boson produced in association with jets in pp collisions in the recent LHC data from ATLAS and CMS, without the need of an unexpectedly hard intrinsic Gaussian distribution with an rms value of PTRMS = 2.2~GeV in parton's wave function. In our view, this can be interpreted as providing a rigorous basis for the phenomenological correctness of such unexpectedly hard distributions insofar as describing
these data using the usual unimproved DGLAP-CS showers is concerned. 
\clearpage
\appendix
\section{Scale Factors for Theoretical Predictions}
\begin{table}[h!]
\centering
\begin{tabular}{ccccc } 
 \hline
 Figure number & $\alpha_{\texttt{HERWIRI}}$ & $\alpha_{\texttt{HERWIG}}$& {\Large\strut}$\big(\frac{\chi^2}{d.o.f}\big)_{\texttt{HERWIRI}}$&$\big(\frac{\chi^2}{d.o.f}\big)_{\texttt{HERWIG}}$\\ [0.8ex] 
 \hline
Figure. 1 & ~~0.0201~~ & ~~0.02023~~& 0.76& 2.04 \\
Figure. 2 & 0.0155 & 0.015 & 1.13 & 0.96 \\

 Figure. 3 & 0.03113 & 0.03241 & 1.19 & 1.49 \\

 Figure. 4& 0.03501 & 0.03221 & 1.06 & 1.69  \\
 
 Figure. 5& 0.01460 & 0.01481 & 0.27 & 0.20 \\ 

 Figure. 6& 0.01562 & 0.01141 & 3.27 & 3.96  \\

 Figure. 7& 0.03978 & 0.04038 & 0.35 & 0.71 \\

 Figure. 8& 0.05890 & 0.06062 & 1.01 & 0.63 \\

 Figure. 9& 0.02850 & 0.03601 & 1.05 & 0.43 \\

 Figure. 10& 0.01311 & 0.0128 & 1.18 & 1.69 \\

 Figure. 11& 0.08608 & 0.08051 & 2.08 & 4.77 \\

 Figure. 12& 0.01311 & 0.01324 & 1.59 & 0.78 \\
 
 Figure. 13& 0.01322 & 0.01328 & 1.46 & 0.49 \\

 Figure. 14& 0.01980 & 0.01920 & 0.59 & 0.96 \\

 Figure. 15& 0.01521 & 0.0139 & 2.50 & 0.76 \\
 
 Figure. 16& 0.03116 & 0.03012 & 2.25 & 1.26 \\

 Figure. 17& 0.03301 & 0.03178 & 2.36 & 1.09 \\

 Figure. 18& 0.01476 & 0.01073 & 2.71 & 2.01 \\

 Figure. 19& 0.01318 & 0.01231 & 3.73 & 0.80 \\

 Figure. 20& 0.02013 & 0.02128 & 0.28 & 1.94 \\
 
 Figure. 21& 0.03170 & 0.02913 & 2.96 & 1.65 \\

 Figure. 22& 0.03212 & 0.03091 & 4.39 & 5.27 \\

 Figure. 23& 0.01469 & 0.01108 & 3.80 & 1.05 \\
 
 Figure. 24& 0.01350 & 0.01031 & 4.54 & 1.30 \\

 Figure. 25& 0.5547 & 0.5309 & 4.31 & 0.70 \\

 Figure. 26& 0.5420 & 0.5172 & 7.31 & 1.08\\
 \hline
\end{tabular}
\caption{Summary of the scale factors applied to the theoretical predictions for ATLAS at $\sqrt{s}=7$~TeV. Note that the factor of 2 between the scalings of Figs. 1 and 7 is due to our having simulated for Y instead of the $|Y|$ in the data.}
\label{t3}
\end{table}
\newpage
\section{Scale Factors for CMS at $\sqrt{s}~=~7$~TeV}
\begin{table}[h!]
\centering
\begin{tabular}{ccccc } 
 \hline
 Figure number & $\alpha_{\texttt{HERWIRI}}$ & $\alpha_{\texttt{HERWIG}}$ & {\Large\strut}$\big(\frac{\chi^2}{d.o.f}\big)_{\texttt{HERWIRI}}$&$\big(\frac{\chi^2}{d.o.f}\big)_{\texttt{HERWIG}}$\\  [0.8ex] 
 \hline
 Figure. 27 & ~~0.04373~~ & ~~0.04521~~&~~0.64~~&~~0.35 \\

 Figure. 28 & 0.0615 & 0.061 &~~1.43~~&~~0.73 \\

 Figure. 29 & 0.52852 & 0.4025 &~~2.60~~&~~1.59  \\

 Figure. 30 & 0.04382 & 0.0451 &~~0.57~~&~~0.40  \\

 Figure. 31 & 0.06138 & 0.0599 &~~1.70~~&~~1.36\\ 

 Figure. 32 & 0.5261 & 0.390 &~~4.02~~&~~4.37~~ \\

 Figure. 33 & 0.04635 & 0.046702 &~~0.39~~&~~0.79~~ \\

 Figure. 34 & 0.06175 & 0.062021 &~~1.94~~&~~1.71\\

 Figure. 35 & 0.502 & 0.415 &~~0.82~~&~~0.61~~ \\

 Figure. 36 & 0.0421 & 0.04411 &~~1.26~~&~~2.67~~ \\

 Figure. 37 & 0.06011 & 0.05981 &~~2.73&~~1.48 \\

 Figure. 38 & 0.5212 & 0.3978 &~~0.89~~&~~0.61 \\
 
 Figure. 39 & 0.6836 & 0.559 &~~0.46~~&~~0.56 \\
 
 Figure. 40 & 0.6251 & 0.5551 &~~1.16~~&0.83 \\
 \hline
 \end{tabular}
\caption{Summary of the scale factors applied to the theoretical predictions for CMS at $\sqrt{s}=7$~TeV}
\label{t4}
\end{table}
\newpage
\section{Ratio Plots}
\begin{figure}[h]
    \centering
    \includegraphics[scale=0.4]{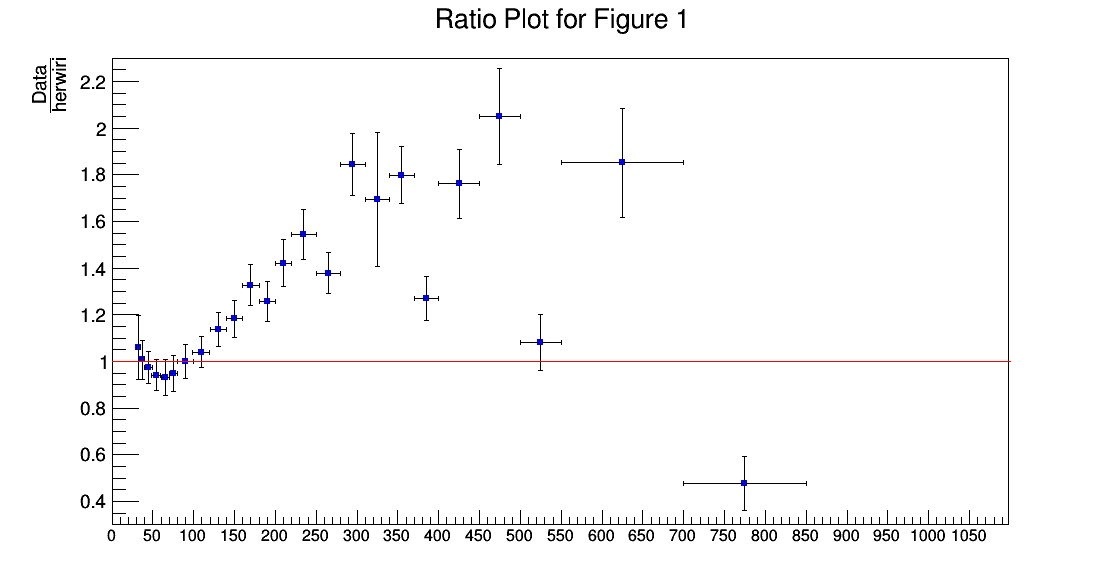}
    \caption{Ratio plot for the production of W~+ jets as a function of the leading-jet $P_{T}$ in $N_{jet}\geq 1$. The data are divided by predictions from MADGRAPH5\_aMC@NLO/HERWIRI1.031.}
    \label{RHERWIRIF1}
\end{figure}
\begin{figure}[H]
    \centering
    \includegraphics[scale=0.4]{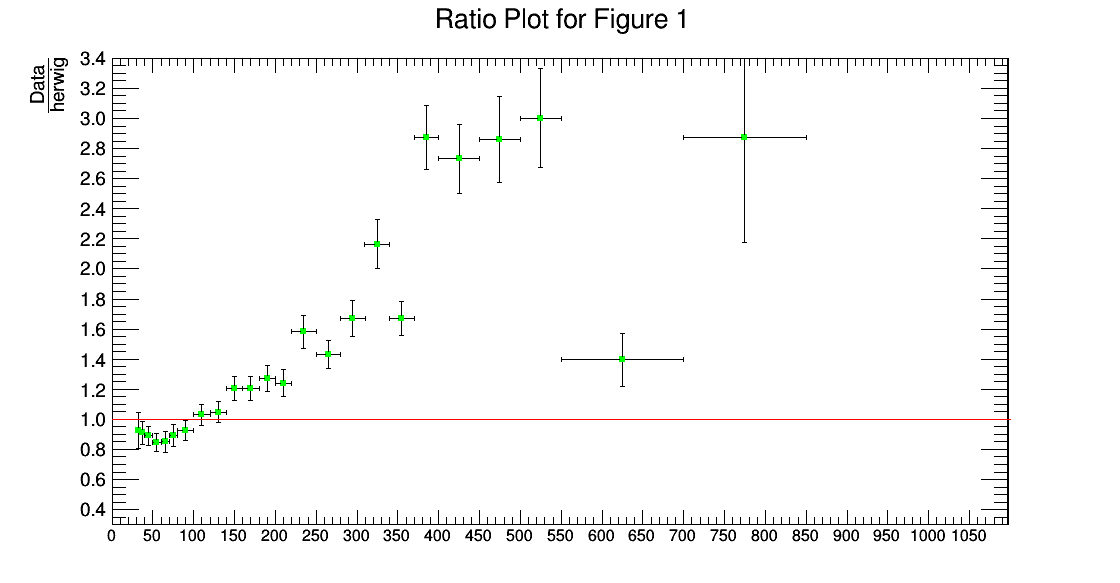}
    \caption{Ratio plot for the production of W~+ jets as a function of the leading-jet $P_{T}$ in $N_{jet}\geq 1$. The data are divided by predictions from MADGRAPH5\_aMC@NLO/HERWIG6.521.}
    \label{RHERWIGF1}
\end{figure}
\begin{figure}[h]
    \centering
    \includegraphics[scale=0.4]{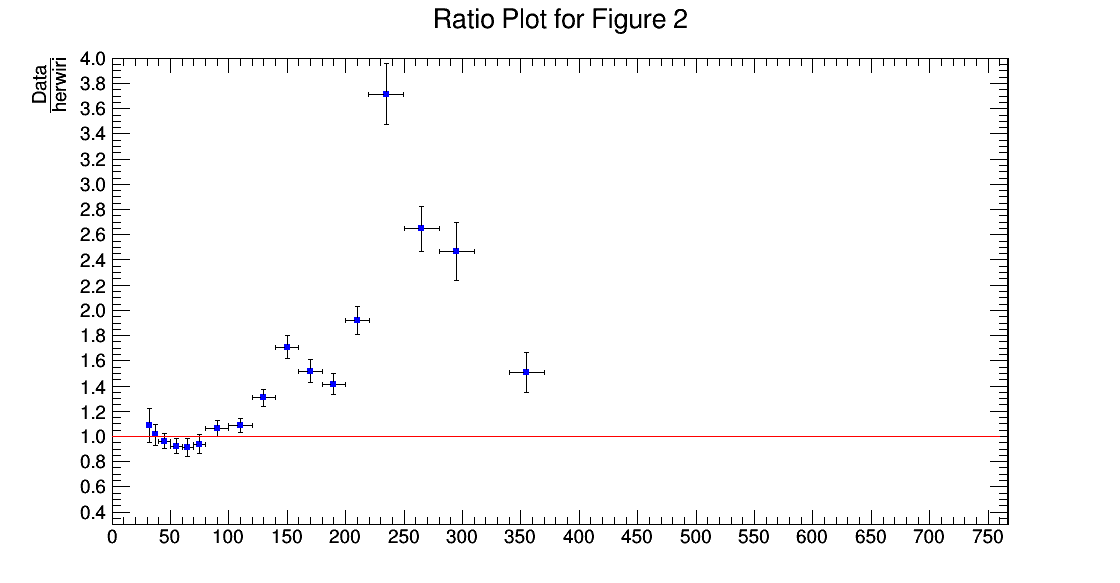}
    \caption{Ratio plot for the production of W~+ jets as a function of the leading-jet $P_{T}$ in $N_{jet}=1$. The data are divided by predictions from MADGRAPH5\_aMC@NLO/HERWIRI1.031.}
    \label{RHERWIRIF2}
\end{figure}
\begin{figure}[h]
    \centering
    \includegraphics[scale=0.4]{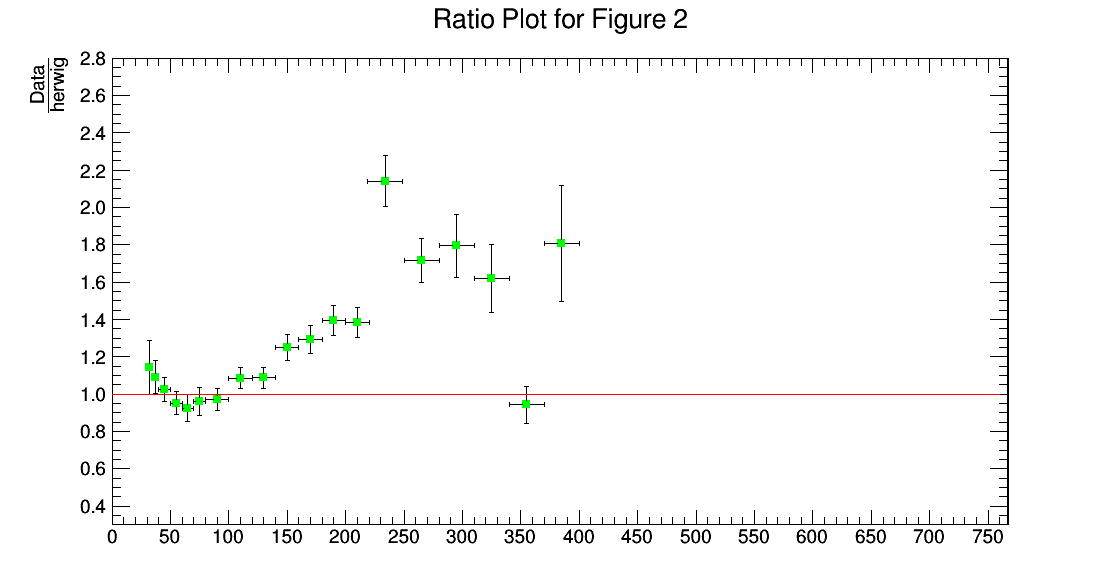}
    \caption{Ratio plot for the production of W~+ jets as a function of the leading-jet $P_{T}$ in $N_{jet}=1$. The data are divided by predictions from MADGRAPH5\_aMC@NLO/HERWIG6.521.}
    \label{RHERWIGF2}
\end{figure}
\begin{figure}[h]
    \centering
    \includegraphics[scale=0.4]{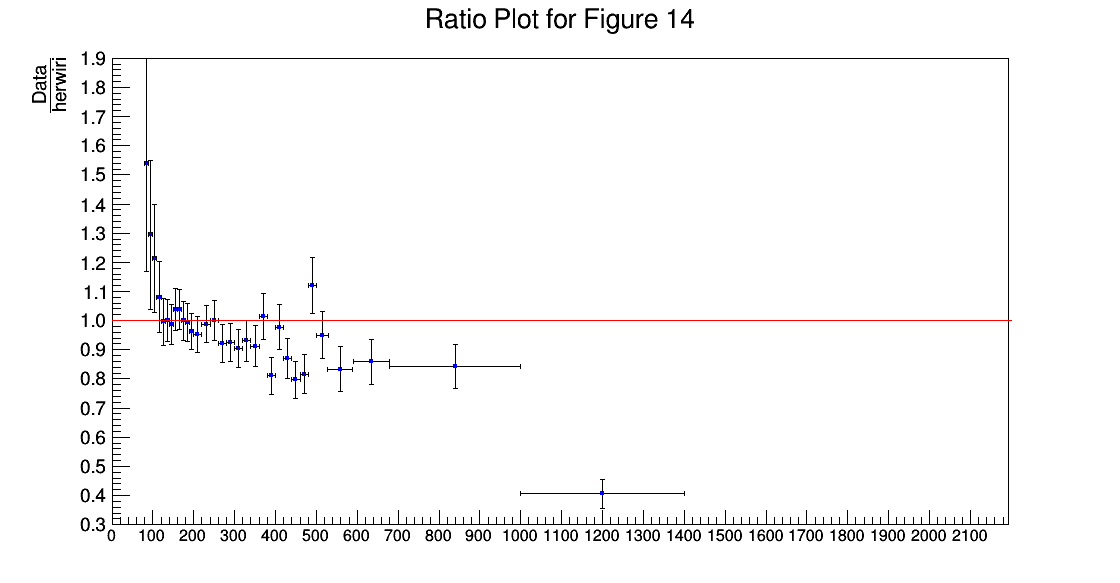}
    \caption{Ratio plot for the production of W~+ jets as a function of the scalar sum $H_{T}$ in $N_{jet}\geq 1.$ The data are divided by predictions from MADGRAPH5\_aMC@NLO/HERWIRI1.031.}
    \label{RHERWIRIF14}
\end{figure}
\begin{figure}[h]
    \centering
    \includegraphics[scale=0.4]{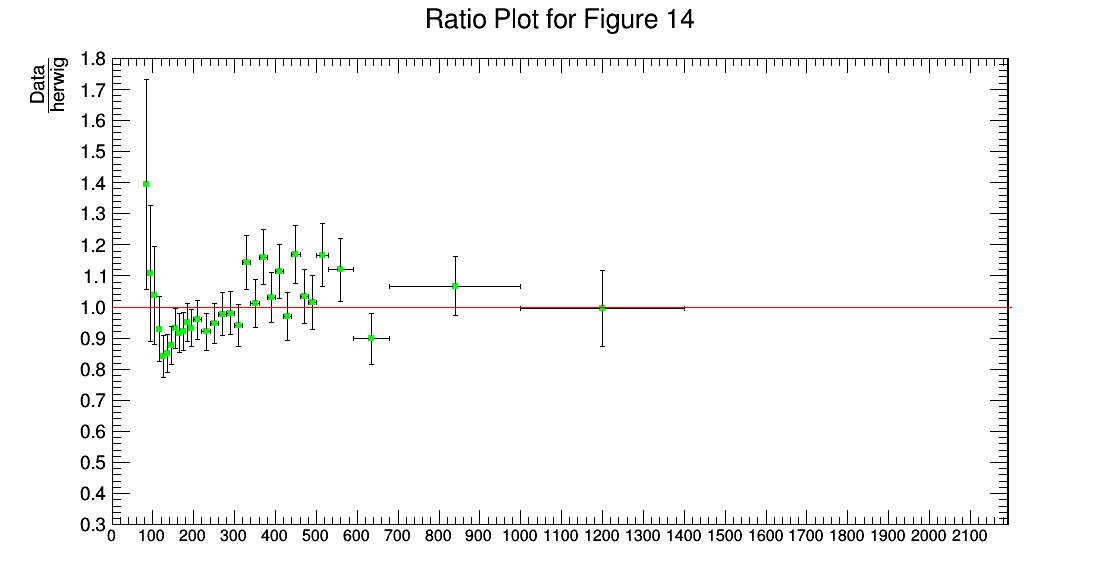}
    \caption{Ratio plot for the production of W~+ jets as a function of the scalar sum $H_{T}$ in $N_{jet}\geq 1.$ The data are divided by predictions from MADGRAPH5\_aMC@NLO/HERWIG6.521.}
    \label{RHERWIGF14}
\end{figure}
\begin{figure}[h]
    \centering
    \includegraphics[scale=0.4]{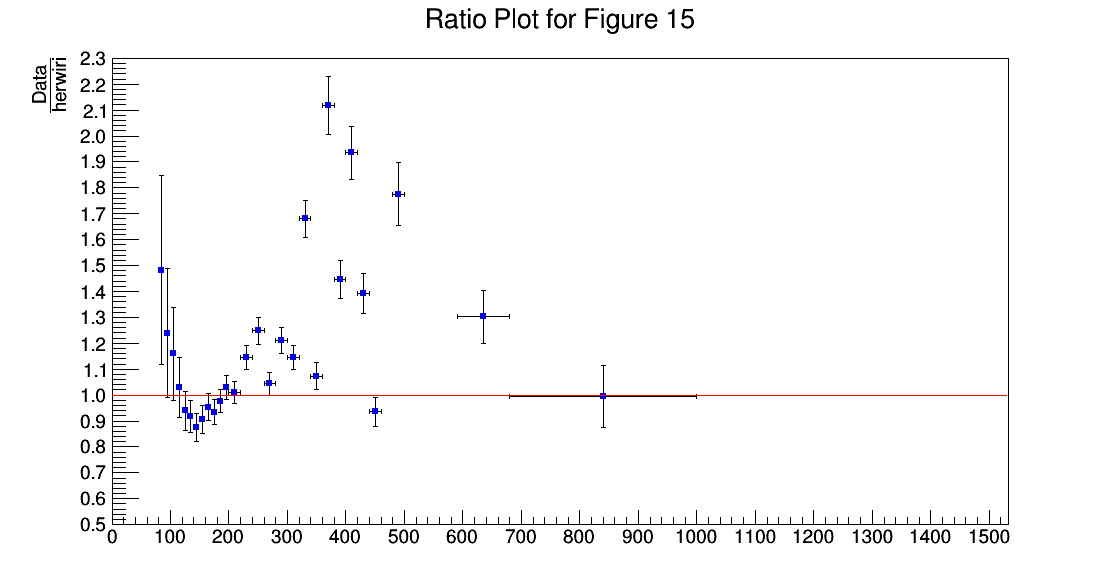}
    \caption{Ratio plot for the production of W~+ jets as a function of the the scalar sum $H_{T}$ in $N_{jet}=1.$ The data are divided by predictions from MADGRAPH5\_aMC@NLO/HERWIRI1.031.}
    \label{RHERWIRIF15}
\end{figure}
\begin{figure}[h]
    \centering
    \includegraphics[scale=0.4]{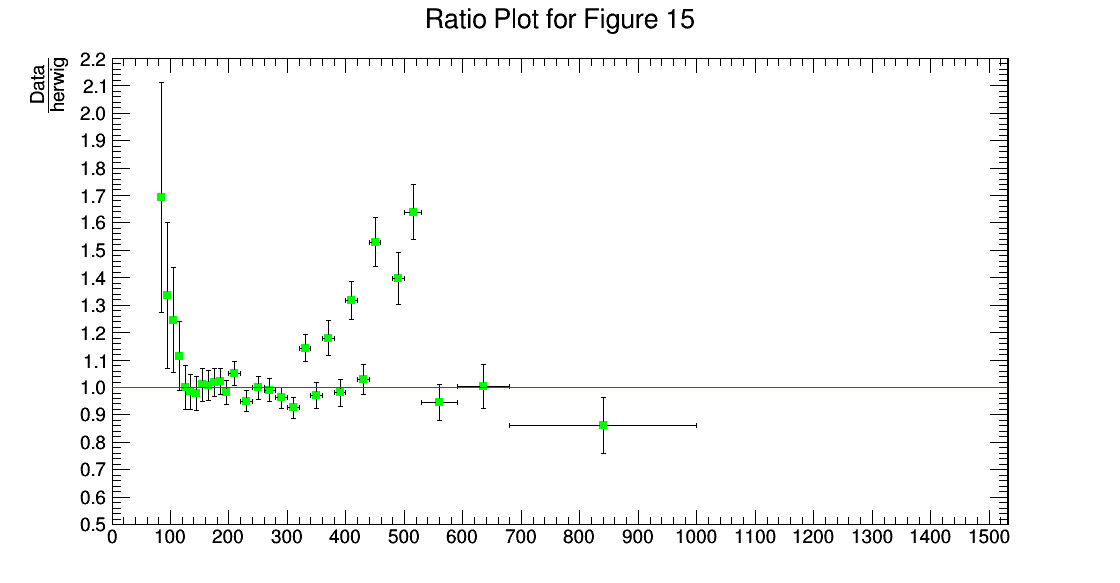}
    \caption{Ratio plot for the production of W~+ jets as a function of the the scalar sum $H_{T}$ in $N_{jet}=1.$ The data are divided by predictions from MADGRAPH5\_aMC@NLO/HERWIG6.521.}
    \label{RHERWIGF15}
\end{figure}

\clearpage
\newpage
\section{Error Plots}
\begin{figure}[h]
\centering
\includegraphics[scale=0.35]{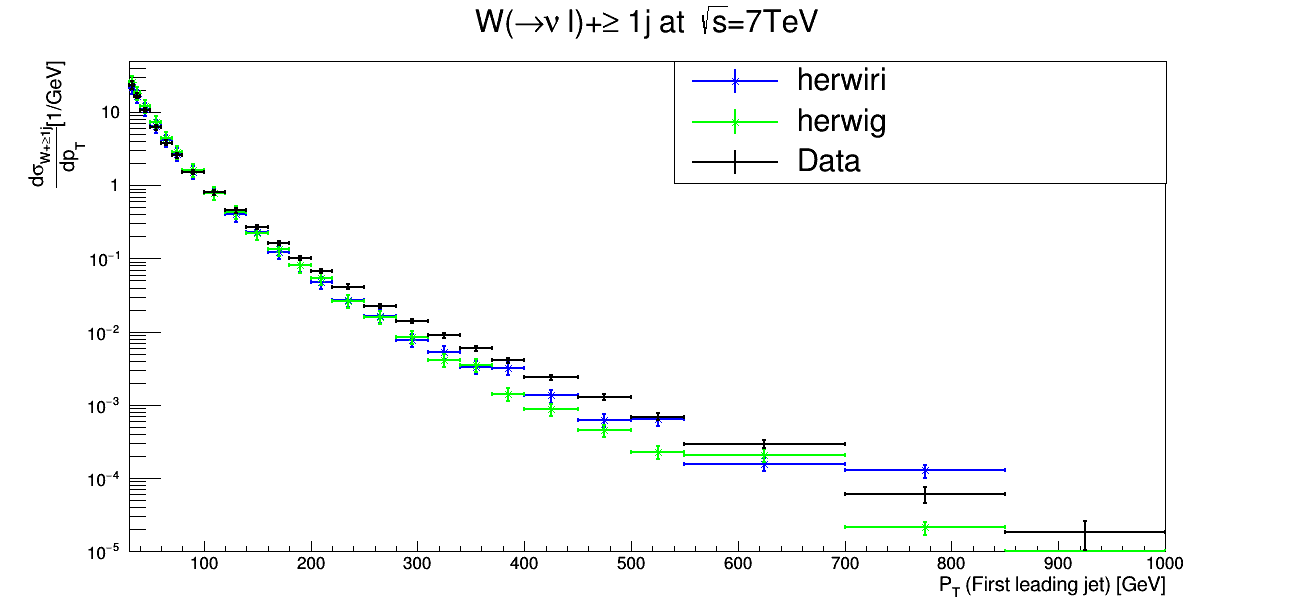}
\caption{Cross section for the production of W~+ jets as a function of the leading-jet $P_{T}$ in $N_{jet}\geq 1$. The data are compared to predictions from MADGRAPH5\_aMC@NLO/HERWIRI1.031 and MADGRAPH5\_aMC@NLO/HERWIG6.521. 20\% theoretical errors are shown for illustration.}
\label{ErrorSample1}
\end{figure}

\begin{figure}[H]
\centering
\includegraphics[scale=0.35]{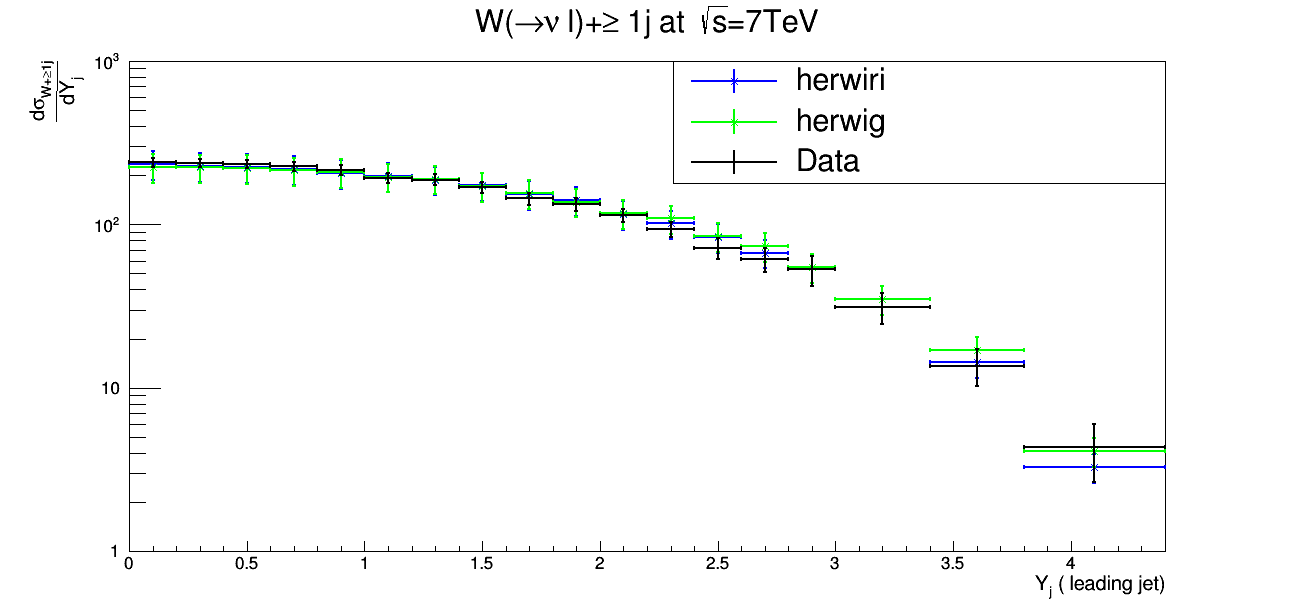}
\caption{Cross section for the production of W~+ jets as a function of the leading-jet $Y_{j}$ in $N_{jet}\geq 1.$ The data are compared to predictions from MADGRAPH5\_aMC@NLO/HERWIRI1.031 and MADGRAPH5\_aMC@NLO/HERWIG6.521. 20\% theoretical errors are shown for illustration.}
\label{ErrorSample2}
\end{figure}
\begin{figure}[H]
\centering
\includegraphics[scale=0.4]{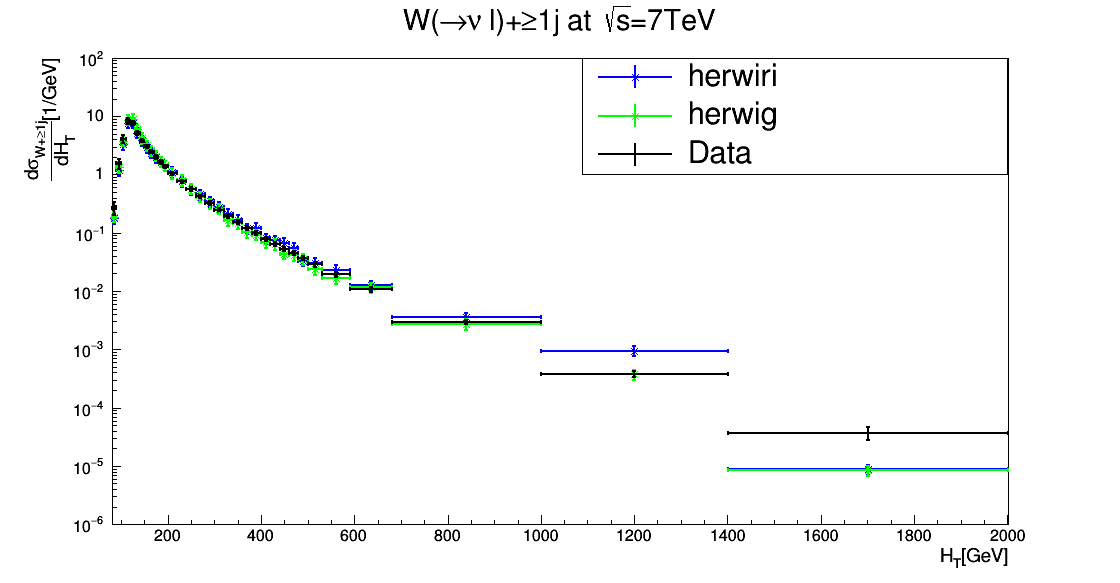}
\caption{Cross section for the production of W~+ jets as a function of the scalar sum $H_{T}$ in $N_{jet}\geq 1.$ The data are compared to predictions from MADGRAPH5\_aMC@NLO/HERWIRI1.031 and MADGRAPH5\_aMC@NLO/HERWIG6.521. 20\% theoretical errors are shown for illustration.}

\end{figure}
\begin{figure}[H]
\centering
\includegraphics[scale=0.4]{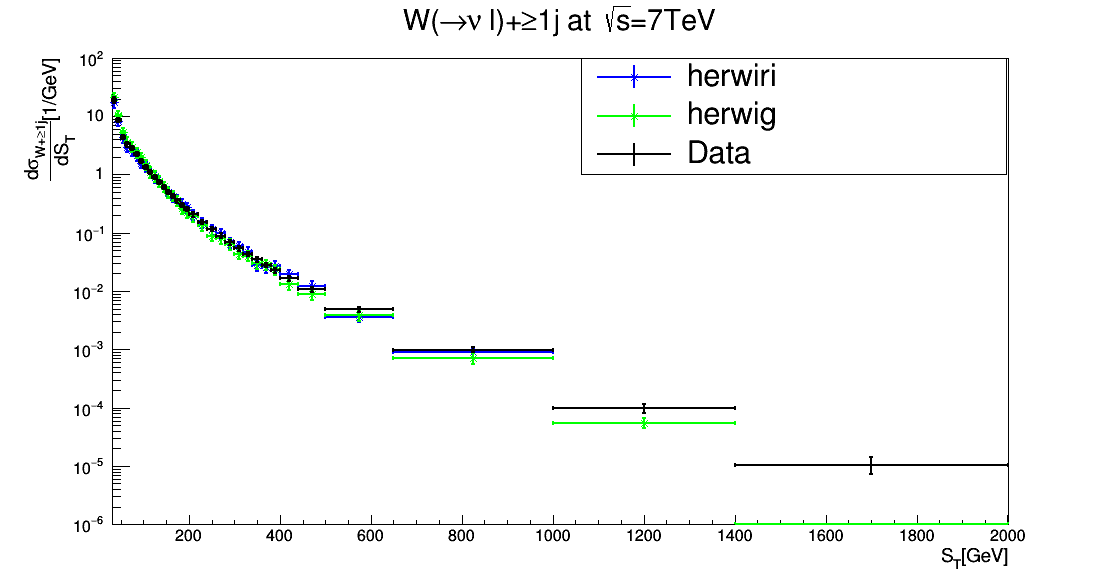}
\caption{Cross section for the production of W~+ jets as a function of the scalar sum $S_{T}$ in $N_{jet}\geq 1.$ The data are compared to predictions from MADGRAPH5\_aMC@NLO/HERWIRI1.031 and MADGRAPH5\_aMC@NLO/HERWIG6.521. 20\% theoretical errors are shown for illustration.}

\end{figure}

\end{document}